\documentclass[%
 reprint,
 superscriptaddress,
 amsmath,amssymb,
 aps,
prd,
]{revtex4-2}

\usepackage{float}
\usepackage{bbold}
\usepackage{braket}
\usepackage{amssymb,amsmath}
\usepackage{tikz}
\usepackage{bbm}
\usepackage{slashed}

\usepackage{amsfonts}
\usepackage{subfigure}
\usepackage{array}
\usepackage{enumerate}

\usepackage{graphicx}
\usepackage{dcolumn}
\usepackage{bm}
\usepackage[dvipsnames]{xcolor}
\usepackage[colorlinks, citecolor=blue,anchorcolor=red,menucolor=red,linkcolor=red,filecolor=red,runcolor=red,urlcolor=blue,frenchlinks=red]{hyperref}
\usepackage[nameinlink]{cleveref}
\crefformat{equation}{Eq.~(#2#1#3)}
\crefformat{figure}{Fig.~#2#1#3}
\crefformat{section}{Sec.~#2#1#3}

\usepackage{ulem}

\graphicspath{{./figures/}{./}}

\begin{document}
\title{Spectral function for pions in magnetic field}
\author{Jie Mei}
\email{meijie22@mails.ucas.ac.cn}
\affiliation{School of Nuclear Science and Technology, University of Chinese Academy of Sciences, Beijing, 100049,  P.R. China}

\author{Rui Wen}
\email{rwen@ucas.ac.cn}
\affiliation{School of Nuclear Science and Technology, University of Chinese Academy of Sciences, Beijing, 100049,  P.R. China}

\author{Min Zhou}
\email{minzhou@stu.xjtu.edu.cn}
\affiliation{School of Physics, Xi'an Jiaotong University, Xi'an, Shaanxi 710049, P.R. China}

\author{Shijun Mao}
\email{maoshijun@mail.xjtu.edu.cn}
\affiliation{School of Physics, Xi'an Jiaotong University, Xi'an, Shaanxi 710049, P.R. China}

\author{Mei Huang}
\email{huangmei@ucas.ac.cn}
\affiliation{School of Nuclear Science and Technology, University of Chinese Academy of Sciences, Beijing, 100049,  P.R. China}
\begin{abstract}
    This study examines the spectral functions of neutral ($\pi_0$) and charged ($\pi_{\pm}$) pions under a uniform magnetic field using the SU(2) Nambu-Jona-Lasinio (NJL) model with the Ritus method. The analysis highlights the complex interplay of magnetic field effects, thermal influences, and chiral symmetry on meson properties in extreme QCD environments. For $\pi_0$, whose properties are governed by the behavior of its constituent quarks, magnetic field-induced Landau levels lead to a multi-peak structure in its spectral function, reflecting stable and resonance solutions that evolve with temperature, showing shifts and critical enhancements near chiral restoration. For $\pi_{\pm}$, cross terms that come from the asymmetry between the constituent quarks introduce Landau cuts alongside Unitary cuts, indicating damping effects, with decay widths narrowing at higher temperatures, suggesting increased stability. 
\end{abstract}
\maketitle

\section{Introduction}
In the early stage of peripheral heavy ion collision, an intense but short-lived magnetic field is created \cite{Preis:2012fh, Gatto:2012sp, DElia:2012ems, Miransky:2015ava, Andersen:2014xxa}. Its intensity can reach up to $1$ to $10 m_{\pi}^2$, which leads to a variety of nonnegligible phenomenon \cite{Kharzeev:2007jp, Gursoy:2014aka, delaIncera:2010wn, Duncan:1992hi,Li:2023tsf}. For non-equilibrium state, one of the most popular subjects is the study of anomalous transport processes, such as chiral magnetic effect (CME)~\cite{fukushima2008chiral} and chiral magnetic wave (CMW)~\cite{Kharzeev:2010gd,Taghavi:2013ena}. For equilibrium state, the effect of a uniform magnetic field on the QCD phase structure is a recently hot topic. For instance, the inverse magnetic catalysis effect (IMC), where the chiral condensates for quarks drop with increasing magnetic field, has been extensively studied for the past decade \cite{Bali:2012edd,Bornyakov:lattice,DElia:lattice,Endrodi:lattice1,Endrodi:lattice2,Bali:2020bcn,Chao:2013qpa,Yu:2014sla,Fayazbakhsh:2014mca,Mei:2020jzn,Xu:2020yag,Lin:2022ied,Kawaguchi:2022dbq,Tavares:2023oln,Chaudhuri:2023djv,Islam:2023zyo,Ferreira:2014kpa, Liu:2016vuw, Avancini:2016fgq, Ayala:2016bbi, Farias:2016gmy, Chen:2021gop, Sheng:2021evj, Mao:2022nfs,Wen:2023qcz, Li:2019nzj, Kamikado:2013pya, Kamikado:2014bua, Fukushima:2012xw, Fu:2017vvg,Bohra:2019ebj,Bohra:2020qom,Wen:2024hgu, Wen:2025cpq}. To fully understand these behaviors, a profound knowledge about the interplay between QCD and electromagnetic field is necessary.

The response of neutral and charged pions on magnetic field is also closely related to the research on phase structure, since they are the (pseudo-) Goldstone bosons under chiral symmetry and their masses' change can be regarded as the signal of chiral symmetry restoration~\cite{Hatsuda:1994pi,Stephanov:2004wx}. When a uniform magnetic field is turned on, the isospin symmetry is broken, making $\pi_{\pm}$ no longer Goldstone bosons~\cite{Fukushima:2012xw,Gatto:2010pt,Miransky:2015ava}. Normally there are two different definitions for meson mass: pole mass and screening mass~\cite{Sheng:2020hge,Mei:2024rjg}. By definition, the pole mass of a particle refers to the mass extracted from the position of the pole in its propagator in momentum space, which corresponds to the physical state of the particle when it is free, whereas the screening mass is associated with the behavior of it in a medium, regarded as the mass ``renormalized'' by the environment~\cite{Bazavov:2019www}. In vacuum where the temperature is zero, pole and screening mass shares the same value.

For screening mass (or pole mass at zero temperature), extensive studies have been performed in both lattice simulation \cite{Bazavov:2019www,Ding:2020hxw,Laudicina:2021gex} and theoretical calculation \cite{Ishii:2013kaa,Czerski:2012fg,Sheng:2020hge, Sheng:2021evj}. For pole mass at finite temperature and magnetic field, however, for now there is only results from effective model calculation, which exhibits a quite peculiar behavior. After including a finite magnetic field, the pole equation starts to possess multiple solutions, and the positions of these solutions are separated by the mass sum of constituent quark mass in different Landau level. As the temperature or chemical potential increases, the physical mass may jump from one solution to another, which is understood as a kind of Mott transition \cite{Mao:2016fha,Mao:2017tcf,Avancini:2018svs,Mao:2019avr,Sheng:2020hge,Mei:2022dkd}. It should be noticed that this behavior can only be observed in quark-level theories like NJL model. Driven by such a strange phenomena, in this work we will explore the spectral function of both neutral and charged pions at finite temperature and magnetic field.

Spectral functions provide a more detailed picture beyond mass spectra. They encode information on the dispersion, width, and possible decay channels of mesonic excitations, and serve as key inputs for evaluating transport coefficients such as shear viscosity~\cite{Tripolt:2013haa,Tripolt:2014wra,Wambach:2014vta,Tripolt:2016cey}. The spectral function of mesons with finite temperature and chemical potential is vastly studied in the past decades, and recently the system under rotation and with chiral imbalance is also taken under consideration \cite{Islam:2018sog,Wei:2021dib,Wei:2023pdf,Ghosh:2023rft}. In hot and magnetized QCD matter, they also offer insights into mass pole, Landau damping~\cite{Bellac:2011kqa}, and critical behavior~\cite{Patkos:2002vr,Patkos:2003cu,Chiku:1997va}. While previous studies have examined vector meson spectral functions~\cite{Sheng:2022ssp,Sheng:2024kgg}, the full analytic structure of pseudoscalar mesons, particularly charged pions, has not been systematically explored.

In this work, we study the spectral functions of both neutral and charged pions at finite temperature in a magnetic field using the SU(2) Nambu–Jona-Lasinio (NJL) model with the Ritus method. For $\pi^0$, we demonstrate how Landau level quantization of the constituent quarks leads to a multi-peak structure in the spectral function. For $\pi^\pm$, we show that cross terms arising from isospin asymmetry in its constituent quarks generate additional Landau cuts, reflecting medium-induced damping processes. Our results highlight the rich analytic structure of mesonic excitations in magnetized media and reveal nontrivial temperature and magnetic field dependence of pion stability and decay width.

The article is arranged as follows: In \cref{sec: method}, we will illustrate how to calculate the mass and spectral function of neutral and charged pions in the frame of SU(2) NJL model in magnetic field; in \cref{sec: Results} we will give the numerical results of the spectral structure for neutral and charged pions separately, and analyse each of the peak structure; in \cref{sec:summary} we will summarize the key findings in this article.
\section{Method}\label{sec: method}
We start with the Lagrangian of SU(2) NJL model in the presence of a uniform magnetic field,
\begin{align}
	\mathcal{L}=\bar{\psi}\left(i\slashed{D}-\hat{m}\right)\psi+G\left[\left(\bar{\psi}\psi\right)^2+\left(\bar{\psi}i\gamma_5\vec{\tau}\psi\right)^2\right].
\end{align}
Here, $\psi$ is two-flavor quark field $\psi=(u,d)^T$, $\hat{m}$ is the quark current mass matrix $\hat{m}=\text{diag}(m_u,m_d)$ with $m_u=m_d=m_0$, which explicitly breaks the chiral symmetry, and $\tau_i$ is $i$-th component of Pauli matrices in flavor space. $G$ is the coupling constant in the scalar and pseudo-scalar channels. The covariant derivative $D_\mu=\partial_\mu-i Q A_\mu$ coupling quarks with electric charge $Q=\text{diag}(q_u,q_d)=\text{diag}(2/3 e,-1/3 e)$ to a gauge field ${\bf B}=\nabla\times{\bf A}$. In this article, we choose the Landau gauge, and the potential $A_{\mu}=(0,0,Bx_1,0)$.

In presence of a uniform magnetic field, the translational invariance of quark in coordinate space is broken, and 4-momentum are no longer ``good" quantum number to the system. Thus if we straightforwardly apply the Fourier transformation to the quark propagator from coordinate space into momentum space, it will introduce a Schwinger phase which compensates for the broken translational and gauge invariance, as discussed in \cite{GomezDumm:2023owj}. A reasonable solution to this situation is to replace the 4-momentum with a new set of ``good" quantum number, which gives rise to the conserved Ritus momentum $\bar{p}=(p_0,0,-s_f\sqrt{2n|q_f B|},p_3)$ with $n$ being the quark Landau level in magnetic field and $s_f=\text{sign}(q_fB)$ the quark sign factor. In this Ritus scheme, the quark propagator with flavor $f$ in coordinate space can be expressed as
\begin{widetext}
\begin{align}
    S_f(x,y)&=\sum_n \int \frac{d^3\tilde{p}}{(2\pi)^3}e^{-i\tilde{p}\cdot(x-y)}P_n(x_1,p_2)D_f(\bar{p})P_n(y_1,p_2)\nonumber\\
    P_n(x_1,p_2)&=\frac{1}{2} \left[g_n^{s_f}(x_1,p_2)+I_ng_{n-1}^{s_f}(x_1,p_2)\right]+\frac{is_f}{2}\left[g_n^{s_f}(x_1,p_2)-I_ng_{n-1}^{s_f}(x_1,p_2)\right]\gamma_1\gamma_2\nonumber\\
    D_f(\bar{p})&=\frac{1}{\gamma\cdot\bar{p}-m_q},
\end{align}
\end{widetext}
where $\tilde{p}=(p_0,0,p_2,p_3)$ is the Fourier transformed momentum and $I_n=1-\delta_{n0}$ is governed by the Landau level. The magnetic field dependent function $g_n^{s_f}(x_1,p_2)=\phi_n(x_1-s_fp_2/|q_fB|)$ is controlled by the Hermite polynomial $H_n(\zeta)$ through
\begin{align}
\phi_n(\zeta)=(2^n n!\sqrt{\pi}|q_fB|^{-1/2})^{-1/2}e^{-\zeta^2|q_fB|/2}H_n(\zeta\sqrt{|q_fB|}) .
\end{align}
Here, $m_q$ is the constituent mass for both $u$ and $d$ quarks, which is determined by chiral condensates with $m_q=m_0-2G\langle\bar{\psi}\psi\rangle$ in NJL model.

In mean field approximation, the thermodynamic potential of the system at finite temperature and under external magnetic field $B$ takes the form of:
\begin{eqnarray}
	\Omega_{\text{MF}}(T,B)=\frac{(m_q-m_0)^2}{4G}+\Omega_q(T,B),
\end{eqnarray}
with the contribution from quarks:
\begin{eqnarray}
	\Omega_{q}(T,B)=\text{Tr}_{\{c,f,s,x\}}\ln\left(\frac{1}{T}S^{-1}(x,x)\right).
\end{eqnarray}
The trace operation is carried out over color (c), flavor (f), spinor (s) degrees of freedom, as well as over the four-dimensional coordinate ($x$). $S^{-1}(x,x)$ is the inverse of quark propagator in coordinate space. After taking the trace in the above spaces, we have
\begin{align}
	\label{omega1}
	\Omega_q &= -N_c \sum_{f=u,d}\frac{|Q_f B|}{2\pi}\sum_{n}\alpha_n \int \frac{d p_3}{2\pi} \Bigg[E_f \nonumber\\
	& +2T\ln\left(1+e^{- \frac{E_f}{T}}\right)\Bigg],
\end{align}
with the quark energy $E_f=\sqrt{p^2_z+2 n |Q_f B|+m_q^2}$ of flavor $f=u,d$, longitudinal momentum $p_3$ and  Landau level $l$, and the degeneracy of Landau levels $\alpha_n=2-\delta_{n0}$.

To determine the effective quark mass, we have to find the ground state by locating the global minimum of the thermodynamic potential
\begin{eqnarray}
	\frac{\partial\Omega_{\text{MF}}(T,B)}{\partial m_q}=0,\nonumber\\
	\frac{\partial^2\Omega_{\text{MF}}(T,B)}{\partial m_q^2}\geq 0.
\end{eqnarray}

In the NJL model, mesons emerge dynamically as collective excitations of quark-antiquark pairs. In the spirit of the random phase approximation (RPA) method, the meson propagator is given by,
\begin{eqnarray}\label{eq:prop}
	U_M(q^2)=\frac{2G}{1-2G\Pi_M(q^2)},
\end{eqnarray}
with polarization function
\begin{align}\label{pimm}
	\Pi_M(x,y)=i\int\!\!\frac{d^4 k}{(2\pi)^4}\text{Tr}_{\{c,f,s\}}\left[\mathcal{V}_{M}^* S(x,y)\mathcal{V}_{M} S(y,x)\right].
\end{align}
The meson vertex is given by
\begin{equation}
	\mathcal{V}_M=\left\{
	\begin{array}{l}
		1,\\
		i\gamma_5\tau_+,\\
		i\gamma_5\tau_-,\\
		i\gamma_5\tau_3,
	\end{array}	\right. 
    \begin{array}{l}
    	M=\sigma\\
    	M=\pi_+\\
    	M=\pi_-\\
    	M=\pi_0,
    \end{array}.
\end{equation}

There is one step missing between \cref{eq:prop} and \cref{pimm}, which is the transformation of polarization function between coordinate and momentum space. At finite magnetic field, this transformation is non-trivial. For neutral mesons which do not interact with the magnetic field as a whole object, the momentum $q = (\omega, \vec{q})$ itself is conserved, and transformation from coordinate space to momentum space is carried out by just the plane wave, which is the solution of Klein-Gordon equation. For charged mesons, though, the transformation kernel from coordinate space to Ritus momentum space, which is the solution of the magnetized Klein-Gordon equation, is
$e^{-i\tilde{p}\cdot x}g_l^{s_M}(x_1,p_2)$ with $l$ being the mesonic Landau level and $\bar{p}$ is conserved.

Spectral functions provide a powerful tool to study the properties of the above mesons, especially in extreme environments like high temperature and intense magnetic field. It encodes the distribution of states with $\omega$ and $p$. 


For mesons, the spectral function is related to the imaginary part of the retarded propagator,
\begin{align}\label{eq:spF}
    \xi(q)&\equiv\frac{1}{\pi}\mathrm{Im}U_M(q)\nonumber\\
    &=\frac{1}{\pi}\frac{4G^2\mathrm{Im}\Pi_M(q)}{(1-2G\mathrm{Re}\Pi_M(q))^2+(2G\mathrm{Im}\Pi_M(q))^2}.
\end{align}
In this work, we set the external momentum $\vec{q} = \vec{0}$ for $\pi_0$ and $q_3 = n = 0$ for $\pi_{\pm}$. This choice focuses on the rest-frame properties and isolates the essential physics of Landau level quantization of constituent quarks while avoiding complications from spatial anisotropy induced by the magnetic field. 
The decay width of a meson can be extracted from its spectral function by measuring the width of peak. In our case, the inclusion of multiple Unitary cuts and Landau cuts (which will be introduced officially in next subsection) will change the shape of the peak structures in spectral function, making their shape deviate from the standard Breit-Wigner structure. An alternative method to calculate the decay width is to replace the meson mass $m_M$ by $m_M-i\Gamma_M/2$, and the pole equation takes the complex form \cite{Mao:2017wmq}
\begin{align}\label{eq:pole}
    1-2G\Pi_M(m_M-i\Gamma_M/2,\mathbf{0})=0.
\end{align}

In practice, the calculation of \cref{eq:pole} usually involves the small-decay-width approximation, which enables us to calculate the mass and decay width separately. The detail of this technique will be shown in the following subsections.

\subsection{$\pi_0$ case}
\begin{figure}[htbp!]
    \includegraphics[width=\columnwidth]{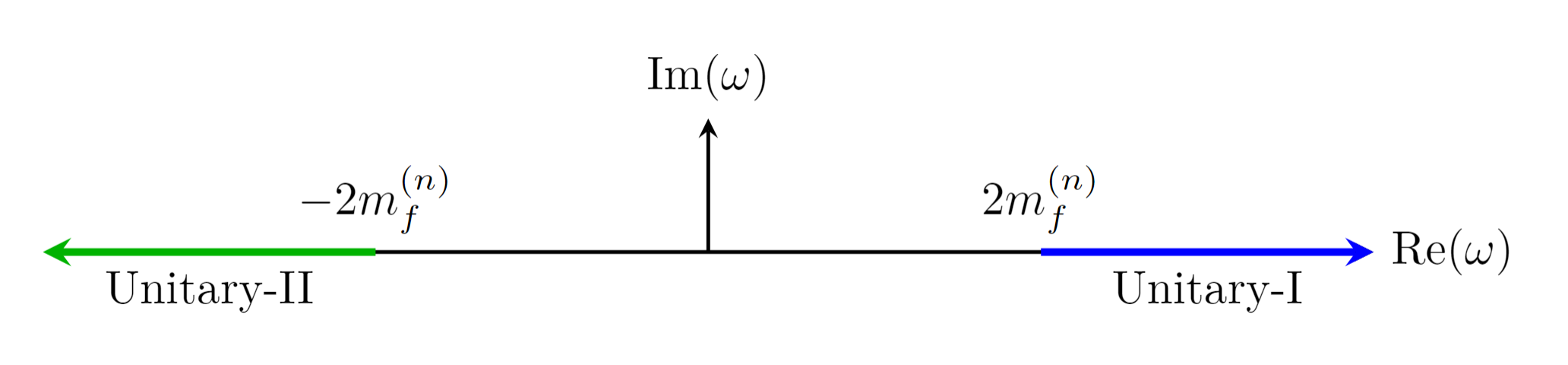}
    \caption{Analytic structure of self-energy $\Pi(\omega,0)$ for $\pi_0$ in complex plane. Here is the Unitary cut at $n$th Landau level of quark with flavor $f$.\label{fig:N_Ucut}}
\end{figure}

For $\pi_0$, the polarization function is given by
\begin{align}
	\Pi_{\pi_0}(\omega)=\sum_{f=u,d}J_1^{(f)}+\omega^2 J_2^{(f)}(\omega)\label{pikaon}
\end{align}
with
\begin{align}
	J_1^{(f)}&=3\!\sum_{n}\alpha_n \frac{|Q_f B|}{2\pi}\!\int \frac{dp_3}{2\pi}\frac{1-2f(E_f)}{E_f}\nonumber\\
	J_2^{(f)}(\omega)&=3\!\sum_{n}\alpha_n \frac{|Q_f B|}{2\pi}\mathcal{P}\!\int \frac{dp_3}{2\pi}\frac{1-2f(E_f)}{E_f(4E_f^2-\omega^2)}\nonumber\\
	&+3i\!\sum_{n}\alpha_n \frac{|Q_f B|}{4\pi}\frac{1-2f(\frac{\omega}{2})}{4\omega\sqrt{\frac{\omega^2}{4}-m_q^2-2n|Q_f B|}}\nonumber\\
    &\times\Theta(\omega-2\sqrt{m_q^2+2n|Q_f B|})\label{kaonj2}	
\end{align}
Here, $\mathcal{P}$ denotes Cauchy principal integral, and $\Theta(k)$ is Heaviside function. Fermi-Dirac distribution function is $f(E)=1/(e^{E/T}+1)$. If we look into the form of the real part of $J_2^{(f)}(\omega)$, we will notice that each time when $\omega-\text{min} E_{f,n}=\omega-2\sqrt{m_q^2+2n|Q_f B|}\rightarrow 0^-$, the function diverges. However, when $\omega-\text{min} E_{f,n}=\omega-2\sqrt{m_q^2+2n|Q_f B|}\rightarrow 0^+$ the function gives a finite value because the principal integral starts to work. This means that the real part of inverse propagator is discontinuous at $\omega-2\sqrt{m_q^2+2n|Q_f B|}\rightarrow 0$. For the imaginary part, similar behaviors can be noticed. It diverges each time when it approaches to the boundary $\omega-2\sqrt{m_q^2+2n|Q_f B|}\rightarrow 0^+$. In another perspective, under a constant magnetic field, $\pi_0$ is composed by a series of one-dimensional quarks with different effective masses $m_f^{(n)}=\sqrt{m_q^2+2n|Q_f B|}$. The one-dimensionality causes the divergence at $\omega\rightarrow 2 m_f^{(n)} + 0^-$, which leads to the possible solution of pole equation for $pi_0$. This feature will cause the pole equation to have multiple solutions of $\omega$. Following the method in \cite{Bellac:2011kqa,Ghosh:2023rft}, branch cuts can be applied at the point of discontinuity and separate the function into several continuous segments. The branch cuts here are referred to as Unitary cut-I and II at $( -\infty,-2m_f^{(n)} ]$ and $[2m_f^{(n)},\infty )$, and the cut structure is shown in \cref{fig:N_Ucut}.

After taking the approximation of small decay width compared with the mass, mass and width in \cref{eq:pole} get decoupled and we can get the mass and width from
\begin{align}
    &1-2G\text{Re}\Pi_{\pi_0}(m_{\pi_0},\mathbf{0})=0,\nonumber\\
    &\Gamma_{\pi_0}\simeq\frac{1}{m_{\pi_0}}\text{Im}\frac{1-2G\sum_{f=u,d}J_1^{(f)}}{2G\sum_{f=u,d}\text{Re}J_2^{(f)}((m_{\pi_0}-i\epsilon)^2)}.\label{eq:polewN}
\end{align}

\subsection{$\pi_{\pm}$ case}

For $\pi_{\pm}$, the cross terms start to appear. The translational invariance for $\pi_\pm$ is broken by a uniform magnetic field, which makes 4-momentum no longer a ``good'' quantum number. Applying the methods adopted in \cite{Mao:2018dqe,GomezDumm:2023owj}, the breaking of translational invariance is absorbed by a Fourier-like transform and the polarization function $\Pi_{\pi_\pm}$ is given by
\begin{widetext}
\begin{align}
\label{Bubblepic}
    &\Pi_{\pi_\pm}(\omega,0,-\sqrt{|eB|},0)=\sum_{f=u,d}J_1^{(f)}+\sum_{n,n'}j_{n,n'}(\omega^2)J_3^{n,n'}(\omega^2),\nonumber\\
    &J_3^{n,n'}(\omega^2)=\mathcal{P}\!\int\frac{dp_3}{2\pi} \frac{1}{4E_nE_{n'}}\left[\frac{f(-E_{n'})-f(E_n)}{\omega+E_n+E_{n'}}+\frac{f(E_{n'})-f(-E_n)}{\omega-E_n-E_{n'}}-\frac{f(E_{n'})-f(E_n)}{\omega-E_n+E_{n'}}-\frac{f(-E_{n'})-f(-E_n)}{\omega+E_n-E_{n'}}\right]\nonumber\\
    &+\frac{i}{2}\times \sum_{p_3^*}\Large\{F_1(p_3^*)\Theta(\omega - E_{n'}(0) - E_n(0)) + \left[F_2(p_3^*)\Theta(n'-2n) + F_3(p_3^*)\Theta(-n'+2n)\right]\nonumber\\
    &\times\Theta(-(\omega^2 - \left[E_{n'}(0) - E_n(0)\right]^2))\Large\},
\end{align}
where the imaginary part is
\begin{align}
    F_1(p_3^*) &= -\frac{1}{4 \omega^2} \frac{f(-E_{n'}(p_3^*)) - f(E_n(p_3^*))}{4 E_n(p_3^*) E_{n'}(p_3^*)}\frac{(\omega + E_n(p_3^*) + E_{n'}(p_3^*))\left[\omega^2 - (E_{n'}(p_3^*) - E_n(p_3^*))^2\right]}{2 p_3^*},\nonumber\\
    F_2(p_3^*) &= \frac{1}{4 \omega^2} \frac{f(-E_{n'}(p_3^*)) - f(-E_n(p_3^*))}{4 E_n(p_3^*) E_{n'}(p_3^*)}\frac{(\omega - E_n(p_3^*) + E_{n'}(p_3^*))\left[\omega^2 - (E_{n'}(p_3^*) + E_n(p_3^*))^2\right]}{2 p_3^*},\nonumber\\
    F_3(p_3^*) &= \frac{1}{4 \omega^2} \frac{f(E_{n'}(p_3^*)) - f(E_n(p_3^*))}{4 E_n(p_3^*) E_{n'}(p_3^*)}\frac{(\omega + E_n(p_3^*) - E_{n'}(p_3^*))\left[\omega^2 - (E_{n'}(p_3^*) + E_n(p_3^*))^2\right]}{2 p_3^*},\nonumber\\
    p_3^* &= \pm\sqrt{\omega^4 + (2 n' |Q_u B| + 2 n |Q_d B|)^2 - 4 \omega^2(m^2 + n' |Q_u B| + n |Q_d B|)} / (2 \omega),
\end{align}
with
\begin{align}
    &j_{n,n'}(\omega^2)=(\frac{\omega^2}{2}-n'|Q_uB|-n|Q_dB|)j_{n,n'}^+-2\sqrt{n'|Q_uB|n|Q_dB|}j_{n,n'}^-,\nonumber\\
    &j_{n,n'}^{\pm}=8N_c\int dx_1dy_1dp_2/(2\pi) g_0^{+}(x_1,k_2)g_0^{+}(y_1,k_2) A_{\pm}(x_1,y_1,k_2),\nonumber\\
    &A_{\pm}(x_1,y_1,k_2)=\alpha_+(x_1,k_2)\alpha_+(y_1,k_2)\pm\alpha_-(x_1,k_2)\alpha_-(y_1,k_2),\nonumber\\
    &\alpha_{\pm}(z,k_2)=\frac{1}{2}[I_ng_{n-1}^{s_d}(z,p_2)g_{n'}^{s_u}(z,q_2)\pm u \leftrightarrow d].
\end{align}
\end{widetext}
Here the quark energies for different flavors are given by $E_{n'}=\sqrt{p_3^2+2n'|Q_uB|+m_q^2}$ and $E_n=\sqrt{p_3^2+2n|Q_dB|+m_q^2}$. In the imaginary part, $p_3^*$ corresponds to the singularities that occur in the momentum integral, and the two Heaviside functions guarantee that the imaginary parts activates only when singularities appear in the integral range. 

\begin{figure}
    \centering
    \includegraphics[width=\columnwidth]{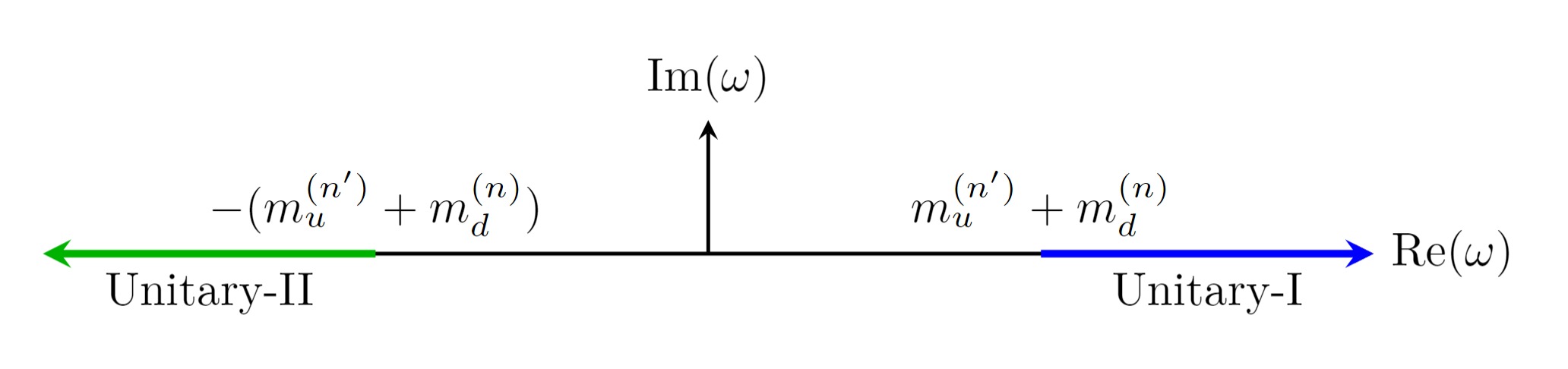}
    \includegraphics[width=\columnwidth]{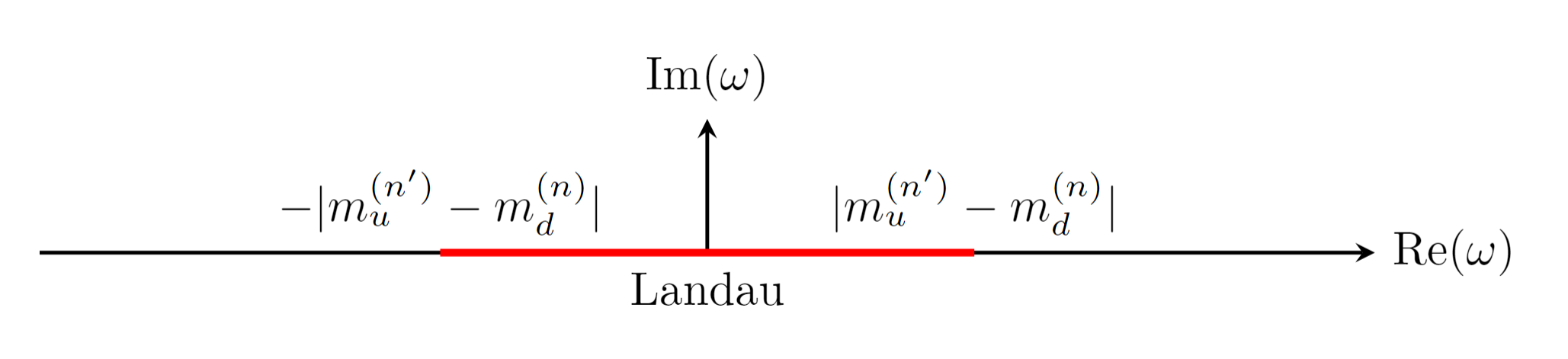}
    \caption{Analytic structure of self-energy $\Pi(\omega,0)$ for $\pi_{\pm}$ in complex plane. Here is the Unitary cut (upper subfigure) and Landau cut (lower subfigure) with $n'$th Landau level of $u$ quark and $n$th Landau level of $d$ quark.}
    \label{fig:C_ULcut}
\end{figure}

Unlike the neutral pion case, if one looks into the polarization function for charged pions, it is noticeable that when a uniform magnetic field is turned on, the orthogonality between the two constituent quarks in different Landau level disappears, and instead there is a coefficient between them named $j_{n,n'}(\omega^2)$. Compared with \cref{kaonj2}, in \cref{Bubblepic} there are two more cross terms (CT), which is the last two terms in $J_3^{n,n'}(\omega^2)$, because of the different quark component between $\pi_0$ and $\pi_{\pm}$. If we look into the polarization function of $\pi_{\pm}$, it is noticed that apart from the discontinuities and divergences at $\omega-\sqrt{m_u^2+2n'|Q_u B|}-\sqrt{m_d^2+2n|Q_d B|}\rightarrow 0$ as is discussed in $\pi_0$ case, more singularities at $\omega-\sqrt{m_u^2+2n'|Q_u B|}+\sqrt{m_d^2+2n|Q_d B|}\rightarrow 0$ and $\omega+\sqrt{m_u^2+2n'|Q_u B|}-\sqrt{m_d^2+2n|Q_d B|}\rightarrow 0$ also can be found, which is referred to as Landau cut. It usually corresponds to the situation of Landau damping, where one particle is emitted from the thermal medium and another is absorbed \cite{Fu:2024rto}. The analytical structures of Unitary cut and Landau cut with $n'$th Landau level of $u$ quark and $n$th Landau level of $d$ quark are separately shown in the upper and lower subfigure of \cref{fig:C_ULcut}.

In practice, the Landau levels of constituent quarks run from 0 to infinite, creating infinite number of branch cuts in complex plane. However, thanks to the suppression of Fermi-Dirac distribution, only a few of them will have visible influence, making it possible to observe a set of continuous lines in the real and imaginary part of inverse propagator, as one can see in the next section.

For $\pi_{\pm}$, we also can take the small-decay width approximation and get the width from
\begin{align}
    &1-2G\text{Re}\Pi_{\pi_{\pm}}(m_{\pi_{\pm}},0,-\sqrt{|eB|},0)=0,\nonumber\\
    &\Gamma_{\pi_{\pm}}\simeq-\frac{2}{m_{\pi_{\pm}}}\nonumber\\
    &\times\text{Im}\frac{1-2G\left[\sum_{f=u,d}J_1^{(f)}-\sum_{n,n'}k_{n,n'}J_3^{n,n'}((m_{\pi_0}-i\epsilon)^2)\right]}{2G\sum_{n,n'}j_{n,n'}^+J_3^{n,n'}((m_{\pi_0}-i\epsilon)^2)},
\end{align}
with
\begin{align}
   & k_{n,n'}\nonumber\\
   &=(n'|Q_uB|+n|Q_dB|)j_{n,n'}^++2\sqrt{n'|Q_uB|n|Q_dB|}j_{n,n'}^-.
\end{align}

\section{Numerical Results}\label{sec: Results}

In this work, we apply the gauge invariant Pauli-Villars regularization, and by fitting the physical quantities, i.e., the chiral condensate $\langle\bar{\psi}\psi\rangle=(-0.25\text{GeV})^3$ and pion decay constant $f_{\pi}=0.093\text{GeV}$ in vacuum, we fix the current mass of light quarks $m_0=0.005\text{GeV}$, and obtain the coupling constant $G=3.44\text{GeV}^{-2}$ and energy scale $\Lambda=1.127\text{GeV}$.
In the following numerical calculations, we fix the magnetic field strength to $eB = 30m_{\pi}^2$ throughout this study, with $m_{\pi}=0.137\text{GeV}$. 
\subsection{$\pi_0$ case}
\begin{figure*}[htbp!]
		\includegraphics[width=0.66\columnwidth]{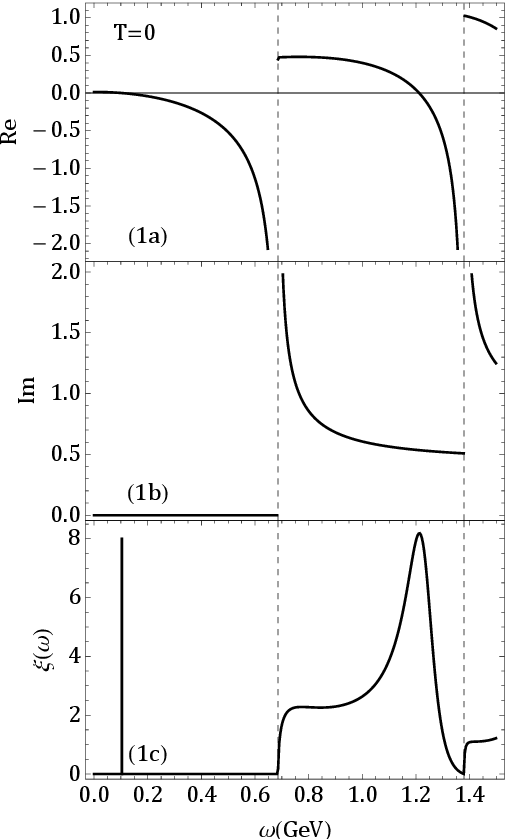}
		\includegraphics[width=0.66\columnwidth]{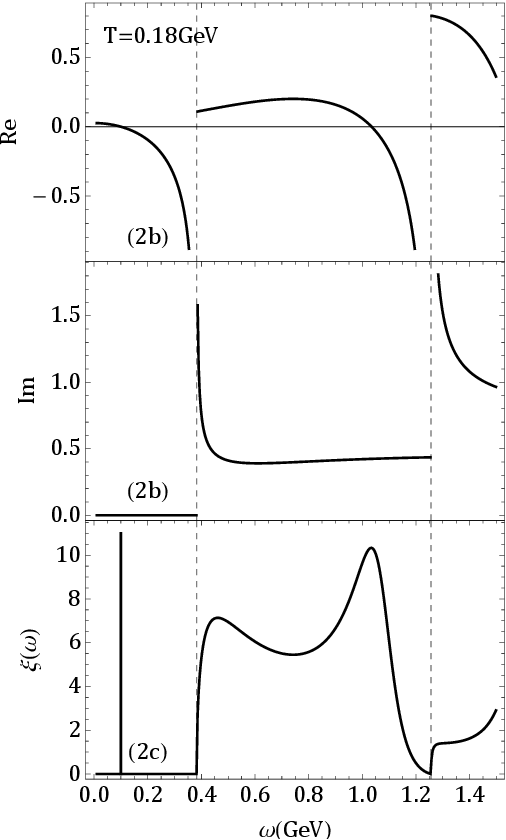}
		\includegraphics[width=0.66\columnwidth]{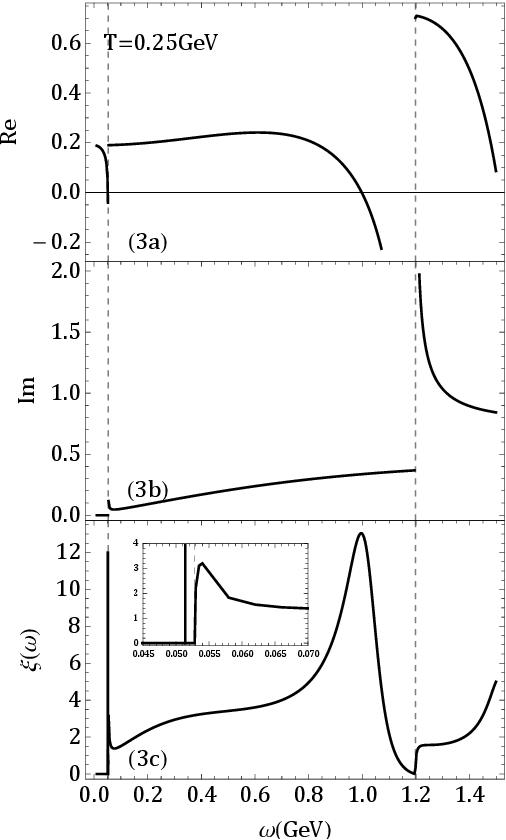}
		\caption{(upper panels) Real part of inverse propagator $1-2G\text{Re}\Pi(\omega)$ as a function of $\omega$; (middle panels) Imaginary part of inverse propagator $2G\text{Im}\Pi(\omega)$ as a function of $\omega$; (lower panels) spectral function of $\pi_0$ as a function of $\omega$. The vertical line represents the delta function, which corresponds to the pole of propagator. The magnetic field strength is set to be $eB=30m_{\pi}^2$, with temperature $T=0$ (first column), $T=0.18\text{GeV}$ (second column), $T=0.25\text{GeV}$ (third column). The vertical dashed lines represents the mass threshold $2m_f^{(n)}$ with $n=0,1,2...$ being the Landau level for constituent quarks.}\label{fig:nT0}
\end{figure*}

With given \cref{eq:spF}, we can start calculating the spectral function of the pions. Here we start with a simpler case with $\pi_0$. In \cref{fig:nT0} we demonstrate the spectral function for $\pi_0$ with different temperatures, as well as the real and imaginary part of the inverse propagator ($1-2G\Pi(\omega)$). In the left column the temperature is set to be zero. (1a) shows $1-2G\text{Re}\Pi(\omega)$ as a function of $\omega$. The curve crosses zero twice within the displayed range, indicating two solutions of the pole equation for the $\pi_0$ mass. The region where the function $1-2G\text{Re}\Pi(\omega)$ diverges correspond to the boundaries where $\omega=2m_q^{(n)}$, with $m_q^{(n)}=\sqrt{m_q^2+2n|q_f B|}$ denoting the effective mass of quark in $n$th Landau level. In (1b), we demonstrate the imaginary part of inverse propagator $2G\text{Im}\Pi(\omega)$. Unlike the case with zero magnetic field where the imaginary part increases continuously around the boundary of $\omega=2m_q$~\cite{Ghosh:2023rft}, in presence of a uniform magnetic field, multiple jumps can be noticed at the boundaries, followed by a fast monotonous decreasing, as a consequence of dimension reduction of constituent quarks. (1c) demonstrates the spectral function of $\pi_0$. In the shown range we can identify two peak structure. One of them is a Dirac delta function, representing the stable particle solution, while the other has a finite width similar to a Breit-Wigner function, denoting the resonance state solution. Unlike in absent magnetic field case, where the spectral function usually possess a standard Breit-Wigner structure, here a ``bump'' at around $0.8\text{GeV}$ can be noticed. Numerically speaking, this structure is believed to result from the fast decline of the imaginary part of the inverse propagator in the vicinity of the discontinuous region. The multi-peak structure of the $\pi_0$ meson spectral function can be understood as follows: although $\pi_0$ is a neutral particle, its constituent quarks, $u$ and $d$, carry electric charge. Their transverse momentum becomes quantized into discrete Landau levels under a magnetic field. Consequently, constituent quarks from different Landau levels may form bound states corresponding to different physical solutions of the Bethe-Salpeter equation. It is generally believed that the first $\delta$-function solution corresponds to the stable excitation of $\pi_0$, whose position is primarily governed by chiral condensate. The subsequent solutions correspond to excited states of the $\pi_0$, with masses affected by both the chiral condensate and the strength of the magnetic field. Another important feature worth noticing is that at left-hand side of the threshold of mass sum of constituent quark, the real part of inverse propagator diverges, while the imaginary part diverges at the right hand side. These two divergences guarantee that the spectral function equals to exact zero at both sides of the threshold, and it is continuous in the whole region.

In the middle column of \cref{fig:nT0}, the temperature is set to $0.18\,\mathrm{GeV}$, where the system's chiral symmetry is still broken, but it is already in the vicinity of the critical temperature $T_c = 0.19\,\mathrm{GeV}$. From the overall trend, (2a) and (2b) panels are very similar to the $T = 0$ case, except for a slight non-monotonic behavior in the curves after the first threshold at $\omega = 0.39\,\mathrm{GeV}$. Taking the imaginary part of the inverse propagator in the middle panel as an example, it decreases sharply just above the threshold, followed by a slower decline and a slight upward turn, until it reaches the jump caused by the next threshold. In (2c), the previously existing delta function shows almost no shift compared with $T=0$ case. Meanwhile, the previously observed bump structure ``grows'' into a new peak. However, this newly emerged peak is not accompanied by a corresponding zero in the real part of the inverse propagator, suggesting that it does not represent a genuine mode excitation.

In the third column, the temperature is set to $T = 0.25\,\mathrm{GeV}$, where the system is in the chiral restoration phase. In (3a), it can be seen that the pole equation's first solution is getting very close to the first threshold at around $\omega\simeq0.05\mathrm{GeV}$, and the solution becomes almost indistinguishable from the threshold. However, the negative divergence at the left side of threshold indicates that as long as the real part of inverse propagator at $\omega=2m_f^{n}$ is positive, there always will be a solution at $(2m_f^{n},2m_f^{n+1})$. In (3b), the non-monotonicity of the imaginary part of the inverse propagator becomes even more obvious. In (3c), after the first threshold in the spectral function,y a rapid decline, forming a sharp edge structure around the threshold $\omega = 2m_q \simeq 0.14\,\mathrm{GeV}$ can be observed. Considering the singular behavior of the imaginary part of the inverse propagator at this point, this peak corresponds to the critical enhancement effect, reflecting the enhancement of the $\pi_0 \rightarrow q\bar{q}$ decay channel.

\begin{figure}[htbp!]
		\includegraphics[width=\columnwidth]{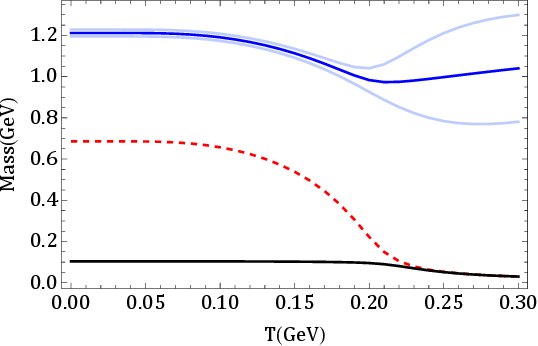}
		\caption{Pole mass of $\pi_0$ as a function of T with magnetic field set to $eB=30m_{\pi}^2$. Two solutions of the pole equation are both shown in the figure (black and blue solid lines). Decay width $\Gamma$ for resonance solution is also shown in the figure by $m_{\pi_0}+\frac{\Gamma}{2}$ and $m_{\pi_0}-\frac{\Gamma}{2}$ (light blue lines). For comparison, the mass sum of the two constituent quark are shown in red dashed line.}\label{fig:nMassT}
\end{figure}

In \cref{fig:nMassT}, we demonstrate the first two solutions of the pole equation for $\pi_0$. The black line stands for the stable particle solution, which correspond to the delta function in spectral function. It can be noticed that this solution shows a non-monotonic behavior: it keeps increasing in the low temperature region, and start to decline at around $T=0.5\text{GeV}$. When temperature gets higher than $T=0.23\text{GeV}$, the first solution for pion masses becomes almost indistinguishable to the threshold $2m_q$. The upper blue line stands for the first resonance solution, and comes with a finite width $\Gamma$ calculated from \cref{eq:polewN}. The mass keep decreasing with temperature at $T<0.21\text{GeV}$ and then starts to increase, showing a non-monotonic behavior. From the decay width, which is stable and small in low temperature region and getting large at around critical temperature, it is evident that this solution remains relatively stable for temperatures below $0.2\,\text{GeV}$, but becomes increasingly unstable beyond this point. 

In the calculation of the meson mass spectrum, previous studies usually only present a single set of physical $\pi_0$ masses, without simultaneously displaying multiple solutions. Therefore, due to different judgment criteria, the reported results vary. In some works, the authors consider the stable particle solution to correspond to the ground state of $\pi_0$, and thus display the complete results of this stable solution until it is no longer accessible. Afterward, the mass is allowed to jump to the next solution with the lowest mass~\cite{Avancini:2016fgq,Sheng:2020hge}. In another series of works, however, the authors argue that the bound state solution of $\pi_0$ should satisfy the Goldstone theorem, meaning that as the chiral condensate decreases, the mass of the stable particle solution should monotonically increase~\cite{Mao:2018dqe,Mei:2022dkd}. Therefore, any decreasing part of this solution is regarded as un-physical, and the meson mass jumps to the next solution at the maximum value. These two criteria lead to significantly different Mott transition temperatures $T_M$. Using the critical temperature $T_c$ determined by the maximum variation of the order parameter as a reference, $T_M$ obtained by the former criterion is larger than $T_c$, while that from the latter is significantly smaller than $T_c$. In the present work, we have shown in \cref{fig:nT0} and \cref{fig:nMassT} that for $\pi_0$, the stable solution always exists, since the real part of inverse propagator exhibits negative divergent at $\omega=2m$ and positive at $\omega=0$.

\subsection{$\pi_{\pm}$ case}

\begin{figure*}[htbp!]
		\includegraphics[width=0.66\columnwidth]{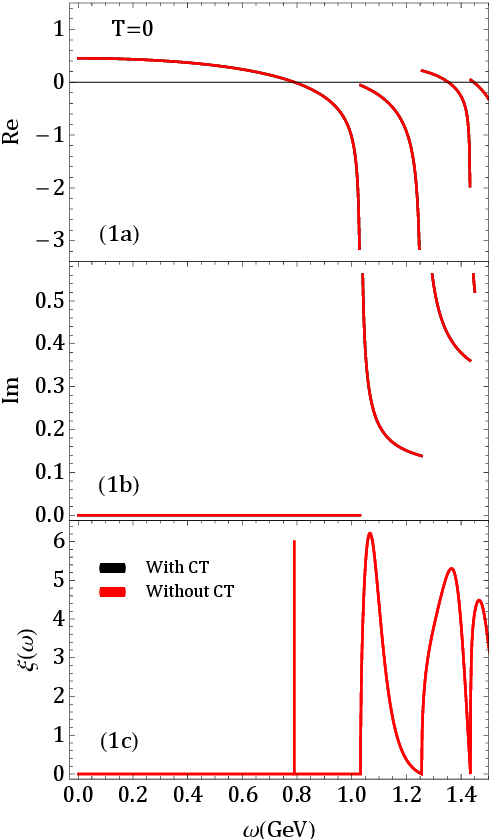}
		\includegraphics[width=0.66\columnwidth]{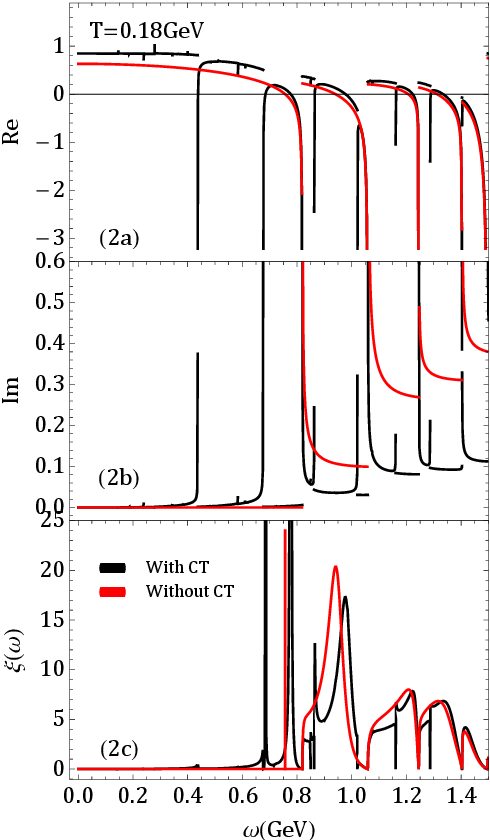}
		\includegraphics[width=0.66\columnwidth]{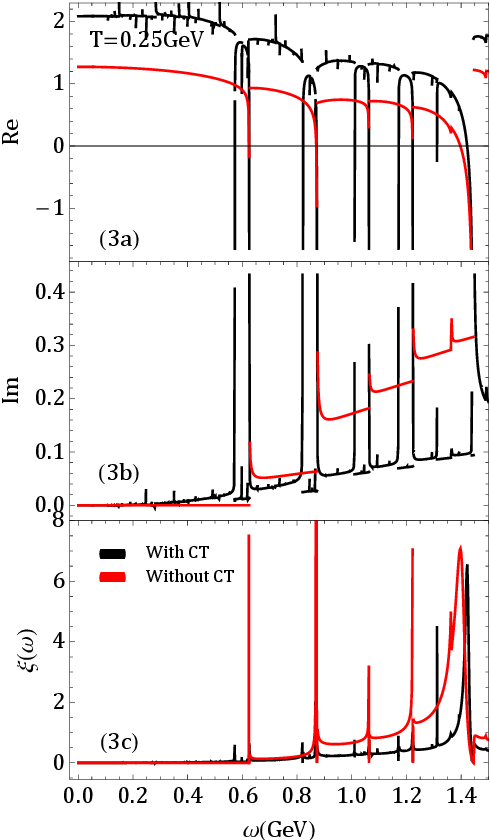}
		\caption{(upper panels) Real part of inverse propagator $1-2G\text{Re}\Pi(\omega)$ for $\pi_{\pm}$ as a function of $\omega$; (middle panels) Imaginary part of inverse propagator $2G\text{Im}\Pi(\omega)$ as a function of $\omega$; (lower panels) spectral function of $\pi_{\pm}$ as a function of $\omega$. The vertical line represents the delta function, which corresponds to the pole of propagator. The magnetic field strength is set to be $eB=30m_{\pi}^2$, with temperature $T=0$ (first column), $T=0.18\text{MeV}$ (second column), $T=0.25\text{GeV}$ (third column). For comparison, the spectral function without considering cross terms is also shown in red color.}\label{fig:cT0}
\end{figure*}

Now we move to results of $\pi_{\pm}$. In \cref{fig:cT0}, we demonstrate the real and imaginary part of the inverse propagator with polarization function defined in \cref{Bubblepic}, as well as the spectral function of $\pi_{\pm}$, with temperature $T=0, 0.18 \text{GeV}, 0.25 \text{GeV}$. To fully understand the effect of cross terms (CT) in \cref{Bubblepic}, we also plot the spectral function without considering CT in red color for comparison. For spectral function, it is observed that CT has no effect at zero temperature since it cancels with itself. As discussed above, CT contributes to the Landau cut, and it appears only in medium and arises from scattering of the external particle with particles in medium, namely the Landau damping effect, as discussed in \cite{Bellac:2011kqa}. At zero temperature, we notice that more discontinuity is found compared with $\pi_0$ case, since there is no orthogonality between the Landau levels of two constituent quarks, and the Unitary cut thresholds occur at $\omega=m_u^{(n')}+m_d^{(n)}$ for $n,n'=0,1,2...$. As for the peak structure, the spectral function possesses four peaks, but the pole equation only has three solutions in the shown range, as one can see in the upper panel. At $\omega=1.07\text{GeV}$, the inverse of the propagator reaches to a local minimum, which is the origin of the peak. 

As the temperature increases to $T=0.18\text{GeV}$, we begin to observe the effects of Landau cuts. For real part in the upper panel, apart from zeros that can be observed in the case without CT, each branch is separated into multiple pieces. At the left-hand side of these CT-induced branch, the results for real part of inverse propagator will become divergent, giving a lot more zeroes. This means that more solutions for the pole equation will be founded. However, as can be seen in (2a), these added solutions all possess a positive slope in $1-2G\text{Re}\Pi(\omega)$. As discussed in \cite{Klevansky:1992qe}, the Yukawa coupling is given by the relationship
\begin{align}
    g_{\pi q\bar{q}}^2=(\partial \Pi/\partial k^2)^{-1}|_{k^2=m_{\pi}^2}.
\end{align}
 The CT-induced solution, which comes with an imaginary Yukawa coupling, can be regarded as an ``unphysical'' mesonic excitation \cite{Ghosh:2023rft}. 
As for the imaginary part, a significant number of new discontinuities also arise, where each continuous segment exhibits a characteristic non-monotonic "U"-shaped profile. 
The Landau cut is located at $\omega=|m_u^{(n')}-m_d^{(n)}|,\ n,n'=0,1,2...$. Unlike the finite number of Unitary cuts within a given energy interval, the number of Landau cuts in a given range is theoretically infinite. However, due to the suppression from the Fermi-Dirac distribution and the parameter $j_{n,n'}(\omega^2)$, only low Landau levels of quarks, particularly those satisfying $n \approx 2n'$, contribute significantly to the Landau cut effect. As the temperature rises, more quark Landau levels need to be taken into account, resulting in an increasing number of discontinuities in high temperature region. For the thresholds created by the Landau cut, the real part of inverse propagator diverges at the right-hand side of the threshold while the imaginary part diverges at the left-hand side. As for spectral function, it also is continuous at the whole range and goes to zero at the threshold.

For the spectral function at finite temperature region, the inclusion of CT starts to possess observable effect. A large number of additional small peak structures appear, which correspond to Landau damping processes. When the temperature further increases to $T=0.25\text{GeV}$, the effects of Landau cuts become even more pronounced, leading to an even greater number of additional peaks in the spectral function. These effects, together with the critical enhancement phenomenon previously discussed for the $\pi_0$ meson, contribute to a rich spectral function structure.

As temperature increasing, the peak sturcture in spectral function keeps getting sharper. Another significant feature at $T=0.25\text{GeV}$ is that, despite the increasing number of resonance states, their widths tend to decrease with rising temperature. It indicates that under a strong magnetic field, these modes of the $\pi_{\pm}$ mesons become more stable at high temperature. As one can see from \cref{eq:spF}, the width of a peak not only relates to the imaginary part of inverse propagator, but also affected by the real part. When the solution of pole equation approaches to the mass threshold $m_u^{(n')}+m_d^{(n)}$, the value of real part changes fast in the vicinity of the solution of pole equation, contributing to a sharp peak structure. 

\begin{figure}[htbp!]
    \centering
    \includegraphics[width=\columnwidth]{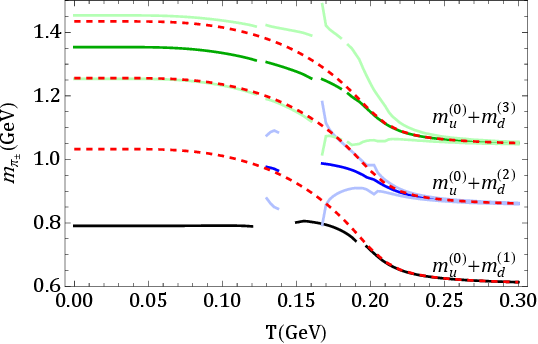}
    \caption{Ground state energy of $\pi_{\pm}$ as a function of T with magnetic field set to $eB=30m_{\pi}^2$. Three lowest solutions of the pole equation are shown in the figure (black, blue and green solid lines). Decay width $\Gamma$ for resonance solution is also shown in the figure by $m_{\pi_{\pm}}+\frac{\Gamma}{2}$ and $m_{\pi_{\pm}}-\frac{\Gamma}{2}$ (light blue and light green lines). For comparison, the mass sum of the two constituent quark in different Landau levels are shown in red dashed lines. }
    \label{fig:cMassT}
\end{figure}

In \cref{fig:cMassT} we show the lowest three physical solutions of the
pole equation for $\pi_{\pm}$ with the corresponding decay width.
The lowest black line, which is the bound state solution,
decreases continuously in the whole temperature region. For the
middle blue line, as one can see in the first row of
\cref{fig:cT0}, the pole equation has no solution for it until
$T>0.14~\text{GeV}$, and all solutions have the same monotonically
decreasing behavior. Its decay width in general gradually becomes
narrower as the temperature increases, consistent with the
analysis on the spectral function in \cref{fig:cT0}. Sudden
changes in the decay width are observed at $T=0.15~\text{GeV}$ and
$T=0.20~\text{GeV}$, which can be attributed to the meson mass
crossing the Landau cut threshold, thereby inducing a nonzero
imaginary part in the polarization function at these temperatures.
Usually only the Landau cut effect with relatively low Landau
levels will have visible influence on the decay width. As for the
upper green line, it has the similar behavior as the bound state
solution. Its width, also like discussed before, generally
decreases with increasing temperature, though several
discontinuous jumps are observed.

It is worth noting that all three solutions exhibit gaps in certain
narrow temperature intervals where no solution can be found. This
phenomenon arises because the Landau cut induces a divergence in
the real part of the inverse propagator precisely at the location
where the zero point originally existed, thereby causing the
disappearance of the solution in these temperature regions.

\section{Summary}\label{sec:summary}
In summary, within the framework of the two-flavor Nambu–Jona-Lasinio (NJL) model, we have systematically investigated the spectral function structures of neutral and charged pions ($\pi_0$ and $\pi_{\pm}$) in the presence of a uniform magnetic field. By employing the Ritus method, we incorporate the breaking of translational invariance into a modified Fourier transformation.

For the neutral pion $\pi_0$ at zero temperature, the spectral function exhibits a distinct multi-peak structure. Some of the peaks correspond to multiple solutions of the pole equation, as indicated by the zeros of the real part of the inverse propagator. The first solution, typically featuring a Dirac delta-function-like peak, represents a stable ground-state excitation, while the subsequent solutions are resonance states associated with finite-width Breit-Wigner-type structures. These multiple peaks arise due to the quantization of transverse momentum into Landau levels for the charged constituent quarks, which leads to a sequence of effective thresholds at $\omega = 2m_q^{(n)} = 2\sqrt{m_q^2 + 2n|Q_f B|}$, separating different excitation modes. However, it should also noticed that not all peak structure is related to a physical particle.

As the temperature increases, the spectral function structure undergoes significant changes. At moderate temperatures below the chiral restoration point, the stable peak slightly shifts non-monotonicaly. Meanwhile, secondary peaks originating from resonances become wider or shift in position. Near and above the chiral critical temperature $T_c$, a sharp enhancement in the spectral function occurs at threshold regions, indicating the onset of critical phenomena like $\pi_0 \rightarrow q\bar{q}$ channel openings. As the temperature gets higher, the mass of the stable $\pi_0$ solution decreases and approaches the quark-antiquark threshold, making it challenging to distinguish from the continuum. The first resonance solution, in contrast, exhibits a non-monotonic temperature dependence and increasing decay width near $T_c$, signaling its diminishing stability.

In comparison, the spectral properties of the charged pions $\pi_{\pm}$ are more intricate. In addition to the Unitary cuts (arising from meson decay into quark-antiquark pairs), their analytic structure also includes Landau cuts due to cross terms in the polarization function. The Landau cuts correspond to thermal scattering processes between mesons and medium constituents, i.e., Landau damping. At zero temperature, these terms vanish due to cancellation, and the results with and without mixing terms coincide. At finite temperatures, however, Landau damping manifests through nontrivial structures in the imaginary part of the inverse propagator, especially a characteristic ``U''-shaped non-monotonicity after each threshold, and a proliferation of small peaks in the spectral function.

The Unitary cuts appear at $\omega = m_u^{(n')} + m_d^{(n)}$ for $n, n' = 0,1,2,\dots$, and increase monotonically with Landau level index, yielding only a finite number of thresholds within a given energy window. By contrast, Landau cuts occur at $\omega = |m_u^{(n')} - m_d^{(n)}|$, potentially leading to an infinite number of discontinuities. However, due to the Fermi-Dirac suppression and the structure of the overlap integral $j_{n,n'}(\omega^2)$, only a limited subset---primarily from low Landau levels with $n \approx 2n'$---exert observable influence.

Regarding the meson mass spectrum, all solutions for $\pi_{\pm}$ exhibit a monotonic decrease with increasing temperature under strong magnetic fields. Their decay widths display sudden threshold-induced jumps when the meson mass crosses Landau cut boundaries. Aside from these threshold effects, the widths generally decrease with temperature, implying that the resonance modes of $\pi_{\pm}$ become increasingly stable at high temperatures, in contrast to the expected thermal broadening in the absence of a magnetic field.

The spectral functions obtained in this work can provide input for understanding various transport phenomena and response functions in magnetized QCD matter. In particular, the detailed analytic structure—including the positions and widths of peaks, as well as the threshold behaviors associated with Unitary and Landau cuts—directly determines the dilepton production rate, photon emission, and thermal medium modifications of hadron properties. Future extensions of this work could employ these spectral functions to calculate transport coefficients such as shear viscosity, electrical conductivity, and thermal conductivity in the presence of strong magnetic fields. Additionally, the multi-peak structure and temperature evolution of the spectral functions may have observable consequences in heavy-ion collision experiments, where the system traverses different temperature regimes. 

\begin{acknowledgments}
In this work, Mei Huang is supported in part by the National Natural Science Foundation of China (NSFC) Grant Nos: 12235016, 12221005, and Shijun Mao is supported by the National Natural Science Foundation of China (NSFC) Grant Nos: 12275204.
\end{acknowledgments}

\bibliography{apssamp}

\begin{thebibliography}{81}%
\makeatletter
\providecommand \@ifxundefined [1]{%
 \@ifx{#1\undefined}
}%
\providecommand \@ifnum [1]{%
 \ifnum #1\expandafter \@firstoftwo
 \else \expandafter \@secondoftwo
 \fi
}%
\providecommand \@ifx [1]{%
 \ifx #1\expandafter \@firstoftwo
 \else \expandafter \@secondoftwo
 \fi
}%
\providecommand \natexlab [1]{#1}%
\providecommand \enquote  [1]{``#1''}%
\providecommand \bibnamefont  [1]{#1}%
\providecommand \bibfnamefont [1]{#1}%
\providecommand \citenamefont [1]{#1}%
\providecommand \href@noop [0]{\@secondoftwo}%
\providecommand \href [0]{\begingroup \@sanitize@url \@href}%
\providecommand \@href[1]{\@@startlink{#1}\@@href}%
\providecommand \@@href[1]{\endgroup#1\@@endlink}%
\providecommand \@sanitize@url [0]{\catcode `\\12\catcode `\$12\catcode
  `\&12\catcode `\#12\catcode `\^12\catcode `\_12\catcode `\%12\relax}%
\providecommand \@@startlink[1]{}%
\providecommand \@@endlink[0]{}%
\providecommand \url  [0]{\begingroup\@sanitize@url \@url }%
\providecommand \@url [1]{\endgroup\@href {#1}{\urlprefix }}%
\providecommand \urlprefix  [0]{URL }%
\providecommand \Eprint [0]{\href }%
\providecommand \doibase [0]{https://doi.org/}%
\providecommand \selectlanguage [0]{\@gobble}%
\providecommand \bibinfo  [0]{\@secondoftwo}%
\providecommand \bibfield  [0]{\@secondoftwo}%
\providecommand \translation [1]{[#1]}%
\providecommand \BibitemOpen [0]{}%
\providecommand \bibitemStop [0]{}%
\providecommand \bibitemNoStop [0]{.\EOS\space}%
\providecommand \EOS [0]{\spacefactor3000\relax}%
\providecommand \BibitemShut  [1]{\csname bibitem#1\endcsname}%
\let\auto@bib@innerbib\@empty
\bibitem [{\citenamefont {Preis}\ \emph {et~al.}(2013)\citenamefont {Preis},
  \citenamefont {Rebhan},\ and\ \citenamefont {Schmitt}}]{Preis:2012fh}%
  \BibitemOpen
  \bibfield  {author} {\bibinfo {author} {\bibfnamefont {F.}~\bibnamefont
  {Preis}}, \bibinfo {author} {\bibfnamefont {A.}~\bibnamefont {Rebhan}},\ and\
  \bibinfo {author} {\bibfnamefont {A.}~\bibnamefont {Schmitt}},\ }\bibfield
  {title} {\bibinfo {title} {{Inverse magnetic catalysis in field theory and
  gauge-gravity duality}},\ }\href
  {https://doi.org/10.1007/978-3-642-37305-3_3} {\bibfield  {journal} {\bibinfo
   {journal} {Lect. Notes Phys.}\ }\textbf {\bibinfo {volume} {871}},\ \bibinfo
  {pages} {51} (\bibinfo {year} {2013})},\ \Eprint
  {https://arxiv.org/abs/1208.0536} {arXiv:1208.0536 [hep-ph]} \BibitemShut
  {NoStop}%
\bibitem [{\citenamefont {Gatto}\ and\ \citenamefont
  {Ruggieri}(2013)}]{Gatto:2012sp}%
  \BibitemOpen
  \bibfield  {author} {\bibinfo {author} {\bibfnamefont {R.}~\bibnamefont
  {Gatto}}\ and\ \bibinfo {author} {\bibfnamefont {M.}~\bibnamefont
  {Ruggieri}},\ }\bibfield  {title} {\bibinfo {title} {{Quark Matter in a
  Strong Magnetic Background}},\ }\href
  {https://doi.org/10.1007/978-3-642-37305-3_4} {\bibfield  {journal} {\bibinfo
   {journal} {Lect. Notes Phys.}\ }\textbf {\bibinfo {volume} {871}},\ \bibinfo
  {pages} {87} (\bibinfo {year} {2013})},\ \Eprint
  {https://arxiv.org/abs/1207.3190} {arXiv:1207.3190 [hep-ph]} \BibitemShut
  {NoStop}%
\bibitem [{\citenamefont {D'Elia}(2013)}]{DElia:2012ems}%
  \BibitemOpen
  \bibfield  {author} {\bibinfo {author} {\bibfnamefont {M.}~\bibnamefont
  {D'Elia}},\ }\bibfield  {title} {\bibinfo {title} {{Lattice QCD Simulations
  in External Background Fields}},\ }\href
  {https://doi.org/10.1007/978-3-642-37305-3_7} {\bibfield  {journal} {\bibinfo
   {journal} {Lect. Notes Phys.}\ }\textbf {\bibinfo {volume} {871}},\ \bibinfo
  {pages} {181} (\bibinfo {year} {2013})},\ \Eprint
  {https://arxiv.org/abs/1209.0374} {arXiv:1209.0374 [hep-lat]} \BibitemShut
  {NoStop}%
\bibitem [{\citenamefont {Miransky}\ and\ \citenamefont
  {Shovkovy}(2015)}]{Miransky:2015ava}%
  \BibitemOpen
  \bibfield  {author} {\bibinfo {author} {\bibfnamefont {V.~A.}\ \bibnamefont
  {Miransky}}\ and\ \bibinfo {author} {\bibfnamefont {I.~A.}\ \bibnamefont
  {Shovkovy}},\ }\bibfield  {title} {\bibinfo {title} {{Quantum field theory in
  a magnetic field: From quantum chromodynamics to graphene and Dirac
  semimetals}},\ }\href {https://doi.org/10.1016/j.physrep.2015.02.003}
  {\bibfield  {journal} {\bibinfo  {journal} {Phys. Rept.}\ }\textbf {\bibinfo
  {volume} {576}},\ \bibinfo {pages} {1} (\bibinfo {year} {2015})},\ \Eprint
  {https://arxiv.org/abs/1503.00732} {arXiv:1503.00732 [hep-ph]} \BibitemShut
  {NoStop}%
\bibitem [{\citenamefont {Andersen}\ \emph {et~al.}(2016)\citenamefont
  {Andersen}, \citenamefont {Naylor},\ and\ \citenamefont
  {Tranberg}}]{Andersen:2014xxa}%
  \BibitemOpen
  \bibfield  {author} {\bibinfo {author} {\bibfnamefont {J.~O.}\ \bibnamefont
  {Andersen}}, \bibinfo {author} {\bibfnamefont {W.~R.}\ \bibnamefont
  {Naylor}},\ and\ \bibinfo {author} {\bibfnamefont {A.}~\bibnamefont
  {Tranberg}},\ }\bibfield  {title} {\bibinfo {title} {{Phase diagram of QCD in
  a magnetic field: A review}},\ }\href
  {https://doi.org/10.1103/RevModPhys.88.025001} {\bibfield  {journal}
  {\bibinfo  {journal} {Rev. Mod. Phys.}\ }\textbf {\bibinfo {volume} {88}},\
  \bibinfo {pages} {025001} (\bibinfo {year} {2016})},\ \Eprint
  {https://arxiv.org/abs/1411.7176} {arXiv:1411.7176 [hep-ph]} \BibitemShut
  {NoStop}%
\bibitem [{\citenamefont {Kharzeev}\ \emph {et~al.}(2008)\citenamefont
  {Kharzeev}, \citenamefont {McLerran},\ and\ \citenamefont
  {Warringa}}]{Kharzeev:2007jp}%
  \BibitemOpen
  \bibfield  {author} {\bibinfo {author} {\bibfnamefont {D.~E.}\ \bibnamefont
  {Kharzeev}}, \bibinfo {author} {\bibfnamefont {L.~D.}\ \bibnamefont
  {McLerran}},\ and\ \bibinfo {author} {\bibfnamefont {H.~J.}\ \bibnamefont
  {Warringa}},\ }\bibfield  {title} {\bibinfo {title} {{The Effects of
  topological charge change in heavy ion collisions: 'Event by event P and CP
  violation'}},\ }\href {https://doi.org/10.1016/j.nuclphysa.2008.02.298}
  {\bibfield  {journal} {\bibinfo  {journal} {Nucl. Phys. A}\ }\textbf
  {\bibinfo {volume} {803}},\ \bibinfo {pages} {227} (\bibinfo {year}
  {2008})},\ \Eprint {https://arxiv.org/abs/0711.0950} {arXiv:0711.0950
  [hep-ph]} \BibitemShut {NoStop}%
\bibitem [{\citenamefont {Gursoy}\ \emph {et~al.}(2014)\citenamefont {Gursoy},
  \citenamefont {Kharzeev},\ and\ \citenamefont {Rajagopal}}]{Gursoy:2014aka}%
  \BibitemOpen
  \bibfield  {author} {\bibinfo {author} {\bibfnamefont {U.}~\bibnamefont
  {Gursoy}}, \bibinfo {author} {\bibfnamefont {D.}~\bibnamefont {Kharzeev}},\
  and\ \bibinfo {author} {\bibfnamefont {K.}~\bibnamefont {Rajagopal}},\
  }\bibfield  {title} {\bibinfo {title} {{Magnetohydrodynamics, charged
  currents and directed flow in heavy ion collisions}},\ }\href
  {https://doi.org/10.1103/PhysRevC.89.054905} {\bibfield  {journal} {\bibinfo
  {journal} {Phys. Rev. C}\ }\textbf {\bibinfo {volume} {89}},\ \bibinfo
  {pages} {054905} (\bibinfo {year} {2014})},\ \Eprint
  {https://arxiv.org/abs/1401.3805} {arXiv:1401.3805 [hep-ph]} \BibitemShut
  {NoStop}%
\bibitem [{\citenamefont {de~la Incera}(2011)}]{delaIncera:2010wn}%
  \BibitemOpen
  \bibfield  {author} {\bibinfo {author} {\bibfnamefont {V.}~\bibnamefont
  {de~la Incera}},\ }\bibfield  {title} {\bibinfo {title} {{Nonperturbative
  Physics in a Magnetic Field}},\ }\href {https://doi.org/10.1063/1.3622687}
  {\bibfield  {journal} {\bibinfo  {journal} {AIP Conf. Proc.}\ }\textbf
  {\bibinfo {volume} {1361}},\ \bibinfo {pages} {74} (\bibinfo {year}
  {2011})},\ \Eprint {https://arxiv.org/abs/1004.4931} {arXiv:1004.4931
  [hep-ph]} \BibitemShut {NoStop}%
\bibitem [{\citenamefont {Duncan}\ and\ \citenamefont
  {Thompson}(1992)}]{Duncan:1992hi}%
  \BibitemOpen
  \bibfield  {author} {\bibinfo {author} {\bibfnamefont {R.~C.}\ \bibnamefont
  {Duncan}}\ and\ \bibinfo {author} {\bibfnamefont {C.}~\bibnamefont
  {Thompson}},\ }\bibfield  {title} {\bibinfo {title} {{Formation of very
  strongly magnetized neutron stars - implications for gamma-ray bursts}},\
  }\href {https://doi.org/10.1086/186413} {\bibfield  {journal} {\bibinfo
  {journal} {Astrophys. J. Lett.}\ }\textbf {\bibinfo {volume} {392}},\
  \bibinfo {pages} {L9} (\bibinfo {year} {1992})}\BibitemShut {NoStop}%
\bibitem [{\citenamefont {Li}\ \emph {et~al.}(2023)\citenamefont {Li},
  \citenamefont {Xia}, \citenamefont {Huang},\ and\ \citenamefont
  {Huang}}]{Li:2023tsf}%
  \BibitemOpen
  \bibfield  {author} {\bibinfo {author} {\bibfnamefont {H.}~\bibnamefont
  {Li}}, \bibinfo {author} {\bibfnamefont {X.-L.}\ \bibnamefont {Xia}},
  \bibinfo {author} {\bibfnamefont {X.-G.}\ \bibnamefont {Huang}},\ and\
  \bibinfo {author} {\bibfnamefont {H.~Z.}\ \bibnamefont {Huang}},\ }\bibfield
  {title} {\bibinfo {title} {{Dynamic calculations of magnetic field and
  implications on spin polarization and spin alignment in heavy ion
  collisions}},\ }\href {https://doi.org/10.1103/PhysRevC.108.044902}
  {\bibfield  {journal} {\bibinfo  {journal} {Phys. Rev. C}\ }\textbf {\bibinfo
  {volume} {108}},\ \bibinfo {pages} {044902} (\bibinfo {year} {2023})},\
  \Eprint {https://arxiv.org/abs/2306.02829} {arXiv:2306.02829 [nucl-th]}
  \BibitemShut {NoStop}%
\bibitem [{\citenamefont {Fukushima}\ \emph {et~al.}(2008)\citenamefont
  {Fukushima}, \citenamefont {Kharzeev},\ and\ \citenamefont
  {Warringa}}]{fukushima2008chiral}%
  \BibitemOpen
  \bibfield  {author} {\bibinfo {author} {\bibfnamefont {K.}~\bibnamefont
  {Fukushima}}, \bibinfo {author} {\bibfnamefont {D.~E.}\ \bibnamefont
  {Kharzeev}},\ and\ \bibinfo {author} {\bibfnamefont {H.~J.}\ \bibnamefont
  {Warringa}},\ }\bibfield  {title} {\bibinfo {title} {Chiral magnetic
  effect},\ }\href@noop {} {\bibfield  {journal} {\bibinfo  {journal} {Physical
  Review D—Particles, Fields, Gravitation, and Cosmology}\ }\textbf {\bibinfo
  {volume} {78}},\ \bibinfo {pages} {074033} (\bibinfo {year}
  {2008})}\BibitemShut {NoStop}%
\bibitem [{\citenamefont {Kharzeev}\ and\ \citenamefont
  {Yee}(2011)}]{Kharzeev:2010gd}%
  \BibitemOpen
  \bibfield  {author} {\bibinfo {author} {\bibfnamefont {D.~E.}\ \bibnamefont
  {Kharzeev}}\ and\ \bibinfo {author} {\bibfnamefont {H.-U.}\ \bibnamefont
  {Yee}},\ }\bibfield  {title} {\bibinfo {title} {{Chiral Magnetic Wave}},\
  }\href {https://doi.org/10.1103/PhysRevD.83.085007} {\bibfield  {journal}
  {\bibinfo  {journal} {Phys. Rev. D}\ }\textbf {\bibinfo {volume} {83}},\
  \bibinfo {pages} {085007} (\bibinfo {year} {2011})},\ \Eprint
  {https://arxiv.org/abs/1012.6026} {arXiv:1012.6026 [hep-th]} \BibitemShut
  {NoStop}%
\bibitem [{\citenamefont {Taghavi}\ and\ \citenamefont
  {Wiedemann}(2015)}]{Taghavi:2013ena}%
  \BibitemOpen
  \bibfield  {author} {\bibinfo {author} {\bibfnamefont {S.~F.}\ \bibnamefont
  {Taghavi}}\ and\ \bibinfo {author} {\bibfnamefont {U.~A.}\ \bibnamefont
  {Wiedemann}},\ }\bibfield  {title} {\bibinfo {title} {{Chiral magnetic wave
  in an expanding QCD fluid}},\ }\href
  {https://doi.org/10.1103/PhysRevC.91.024902} {\bibfield  {journal} {\bibinfo
  {journal} {Phys. Rev. C}\ }\textbf {\bibinfo {volume} {91}},\ \bibinfo
  {pages} {024902} (\bibinfo {year} {2015})},\ \Eprint
  {https://arxiv.org/abs/1310.0193} {arXiv:1310.0193 [hep-ph]} \BibitemShut
  {NoStop}%
\bibitem [{\citenamefont {Bali}\ \emph {et~al.}(2012)\citenamefont {Bali} \emph
  {et~al.}}]{Bali:2012edd}%
  \BibitemOpen
  \bibfield  {author} {\bibinfo {author} {\bibfnamefont {G.~S.}\ \bibnamefont
  {Bali}} \emph {et~al.},\ }\bibfield  {title} {\bibinfo {title}
  {{Thermodynamic properties of QCD in external magnetic fields}},\ }\href
  {https://doi.org/10.22323/1.171.0198} {\bibfield  {journal} {\bibinfo
  {journal} {PoS}\ }\textbf {\bibinfo {volume} {ConfinementX}},\ \bibinfo
  {pages} {197} (\bibinfo {year} {2012})},\ \Eprint
  {https://arxiv.org/abs/1301.5826} {arXiv:1301.5826 [hep-lat]} \BibitemShut
  {NoStop}%
\bibitem [{\citenamefont {Bornyakov}\ \emph {et~al.}(2014)\citenamefont
  {Bornyakov}, \citenamefont {Buividovich}, \citenamefont {Cundy},
  \citenamefont {Kochetkov},\ and\ \citenamefont
  {Sch\"afer}}]{Bornyakov:lattice}%
  \BibitemOpen
  \bibfield  {author} {\bibinfo {author} {\bibfnamefont {V.~G.}\ \bibnamefont
  {Bornyakov}}, \bibinfo {author} {\bibfnamefont {P.~V.}\ \bibnamefont
  {Buividovich}}, \bibinfo {author} {\bibfnamefont {N.}~\bibnamefont {Cundy}},
  \bibinfo {author} {\bibfnamefont {O.~A.}\ \bibnamefont {Kochetkov}},\ and\
  \bibinfo {author} {\bibfnamefont {A.}~\bibnamefont {Sch\"afer}},\ }\bibfield
  {title} {\bibinfo {title} {{Deconfinement transition in two-flavor lattice
  QCD with dynamical overlap fermions in an external magnetic field}},\ }\href
  {https://doi.org/10.1103/PhysRevD.90.034501} {\bibfield  {journal} {\bibinfo
  {journal} {Phys. Rev. D}\ }\textbf {\bibinfo {volume} {90}},\ \bibinfo
  {pages} {034501} (\bibinfo {year} {2014})},\ \Eprint
  {https://arxiv.org/abs/1312.5628} {arXiv:1312.5628 [hep-lat]} \BibitemShut
  {NoStop}%
\bibitem [{\citenamefont {D'Elia}\ \emph {et~al.}(2018)\citenamefont {D'Elia},
  \citenamefont {Manigrasso}, \citenamefont {Negro},\ and\ \citenamefont
  {Sanfilippo}}]{DElia:lattice}%
  \BibitemOpen
  \bibfield  {author} {\bibinfo {author} {\bibfnamefont {M.}~\bibnamefont
  {D'Elia}}, \bibinfo {author} {\bibfnamefont {F.}~\bibnamefont {Manigrasso}},
  \bibinfo {author} {\bibfnamefont {F.}~\bibnamefont {Negro}},\ and\ \bibinfo
  {author} {\bibfnamefont {F.}~\bibnamefont {Sanfilippo}},\ }\bibfield  {title}
  {\bibinfo {title} {{QCD phase diagram in a magnetic background for different
  values of the pion mass}},\ }\href
  {https://doi.org/10.1103/PhysRevD.98.054509} {\bibfield  {journal} {\bibinfo
  {journal} {Phys. Rev. D}\ }\textbf {\bibinfo {volume} {98}},\ \bibinfo
  {pages} {054509} (\bibinfo {year} {2018})},\ \Eprint
  {https://arxiv.org/abs/1808.07008} {arXiv:1808.07008 [hep-lat]} \BibitemShut
  {NoStop}%
\bibitem [{\citenamefont {Endrodi}(2015)}]{Endrodi:lattice1}%
  \BibitemOpen
  \bibfield  {author} {\bibinfo {author} {\bibfnamefont {G.}~\bibnamefont
  {Endrodi}},\ }\bibfield  {title} {\bibinfo {title} {{Critical point in the
  QCD phase diagram for extremely strong background magnetic fields}},\ }\href
  {https://doi.org/10.1007/JHEP07(2015)173} {\bibfield  {journal} {\bibinfo
  {journal} {JHEP}\ }\textbf {\bibinfo {volume} {07}},\ \bibinfo {pages}
  {173}},\ \Eprint {https://arxiv.org/abs/1504.08280} {arXiv:1504.08280
  [hep-lat]} \BibitemShut {NoStop}%
\bibitem [{\citenamefont {Endrodi}\ \emph {et~al.}(2019)\citenamefont
  {Endrodi}, \citenamefont {Giordano}, \citenamefont {Katz}, \citenamefont
  {Kov\'acs},\ and\ \citenamefont {Pittler}}]{Endrodi:lattice2}%
  \BibitemOpen
  \bibfield  {author} {\bibinfo {author} {\bibfnamefont {G.}~\bibnamefont
  {Endrodi}}, \bibinfo {author} {\bibfnamefont {M.}~\bibnamefont {Giordano}},
  \bibinfo {author} {\bibfnamefont {S.~D.}\ \bibnamefont {Katz}}, \bibinfo
  {author} {\bibfnamefont {T.~G.}\ \bibnamefont {Kov\'acs}},\ and\ \bibinfo
  {author} {\bibfnamefont {F.}~\bibnamefont {Pittler}},\ }\bibfield  {title}
  {\bibinfo {title} {{Magnetic catalysis and inverse catalysis for heavy
  pions}},\ }\href {https://doi.org/10.1007/JHEP07(2019)007} {\bibfield
  {journal} {\bibinfo  {journal} {JHEP}\ }\textbf {\bibinfo {volume} {07}},\
  \bibinfo {pages} {007}},\ \Eprint {https://arxiv.org/abs/1904.10296}
  {arXiv:1904.10296 [hep-lat]} \BibitemShut {NoStop}%
\bibitem [{\citenamefont {Bali}\ \emph {et~al.}(2020)\citenamefont {Bali},
  \citenamefont {Endr\H{o}di},\ and\ \citenamefont {Piemonte}}]{Bali:2020bcn}%
  \BibitemOpen
  \bibfield  {author} {\bibinfo {author} {\bibfnamefont {G.~S.}\ \bibnamefont
  {Bali}}, \bibinfo {author} {\bibfnamefont {G.}~\bibnamefont {Endr\H{o}di}},\
  and\ \bibinfo {author} {\bibfnamefont {S.}~\bibnamefont {Piemonte}},\
  }\bibfield  {title} {\bibinfo {title} {{Magnetic susceptibility of QCD matter
  and its decomposition from the lattice}},\ }\href
  {https://doi.org/10.1007/JHEP07(2020)183} {\bibfield  {journal} {\bibinfo
  {journal} {JHEP}\ }\textbf {\bibinfo {volume} {07}},\ \bibinfo {pages}
  {183}},\ \Eprint {https://arxiv.org/abs/2004.08778} {arXiv:2004.08778
  [hep-lat]} \BibitemShut {NoStop}%
\bibitem [{\citenamefont {Chao}\ \emph {et~al.}(2013)\citenamefont {Chao},
  \citenamefont {Chu},\ and\ \citenamefont {Huang}}]{Chao:2013qpa}%
  \BibitemOpen
  \bibfield  {author} {\bibinfo {author} {\bibfnamefont {J.}~\bibnamefont
  {Chao}}, \bibinfo {author} {\bibfnamefont {P.}~\bibnamefont {Chu}},\ and\
  \bibinfo {author} {\bibfnamefont {M.}~\bibnamefont {Huang}},\ }\bibfield
  {title} {\bibinfo {title} {{Inverse magnetic catalysis induced by
  sphalerons}},\ }\href {https://doi.org/10.1103/PhysRevD.88.054009} {\bibfield
   {journal} {\bibinfo  {journal} {Phys. Rev. D}\ }\textbf {\bibinfo {volume}
  {88}},\ \bibinfo {pages} {054009} (\bibinfo {year} {2013})},\ \Eprint
  {https://arxiv.org/abs/1305.1100} {arXiv:1305.1100 [hep-ph]} \BibitemShut
  {NoStop}%
\bibitem [{\citenamefont {Yu}\ \emph {et~al.}(2014)\citenamefont {Yu},
  \citenamefont {Liu},\ and\ \citenamefont {Huang}}]{Yu:2014sla}%
  \BibitemOpen
  \bibfield  {author} {\bibinfo {author} {\bibfnamefont {L.}~\bibnamefont
  {Yu}}, \bibinfo {author} {\bibfnamefont {H.}~\bibnamefont {Liu}},\ and\
  \bibinfo {author} {\bibfnamefont {M.}~\bibnamefont {Huang}},\ }\bibfield
  {title} {\bibinfo {title} {{Spontaneous generation of local CP violation and
  inverse magnetic catalysis}},\ }\href
  {https://doi.org/10.1103/PhysRevD.90.074009} {\bibfield  {journal} {\bibinfo
  {journal} {Phys. Rev. D}\ }\textbf {\bibinfo {volume} {90}},\ \bibinfo
  {pages} {074009} (\bibinfo {year} {2014})},\ \Eprint
  {https://arxiv.org/abs/1404.6969} {arXiv:1404.6969 [hep-ph]} \BibitemShut
  {NoStop}%
\bibitem [{\citenamefont {Fayazbakhsh}\ and\ \citenamefont
  {Sadooghi}(2014)}]{Fayazbakhsh:2014mca}%
  \BibitemOpen
  \bibfield  {author} {\bibinfo {author} {\bibfnamefont {S.}~\bibnamefont
  {Fayazbakhsh}}\ and\ \bibinfo {author} {\bibfnamefont {N.}~\bibnamefont
  {Sadooghi}},\ }\bibfield  {title} {\bibinfo {title} {{Anomalous magnetic
  moment of hot quarks, inverse magnetic catalysis, and reentrance of the
  chiral symmetry broken phase}},\ }\href
  {https://doi.org/10.1103/PhysRevD.90.105030} {\bibfield  {journal} {\bibinfo
  {journal} {Phys. Rev. D}\ }\textbf {\bibinfo {volume} {90}},\ \bibinfo
  {pages} {105030} (\bibinfo {year} {2014})},\ \Eprint
  {https://arxiv.org/abs/1408.5457} {arXiv:1408.5457 [hep-ph]} \BibitemShut
  {NoStop}%
\bibitem [{\citenamefont {Mei}\ and\ \citenamefont {Mao}(2020)}]{Mei:2020jzn}%
  \BibitemOpen
  \bibfield  {author} {\bibinfo {author} {\bibfnamefont {J.}~\bibnamefont
  {Mei}}\ and\ \bibinfo {author} {\bibfnamefont {S.}~\bibnamefont {Mao}},\
  }\bibfield  {title} {\bibinfo {title} {{Inverse catalysis effect of the quark
  anomalous magnetic moment to chiral restoration and deconfinement phase
  transitions}},\ }\href {https://doi.org/10.1103/PhysRevD.102.114035}
  {\bibfield  {journal} {\bibinfo  {journal} {Phys. Rev. D}\ }\textbf {\bibinfo
  {volume} {102}},\ \bibinfo {pages} {114035} (\bibinfo {year} {2020})},\
  \Eprint {https://arxiv.org/abs/2008.12123} {arXiv:2008.12123 [hep-ph]}
  \BibitemShut {NoStop}%
\bibitem [{\citenamefont {Xu}\ \emph {et~al.}(2021)\citenamefont {Xu},
  \citenamefont {Chao},\ and\ \citenamefont {Huang}}]{Xu:2020yag}%
  \BibitemOpen
  \bibfield  {author} {\bibinfo {author} {\bibfnamefont {K.}~\bibnamefont
  {Xu}}, \bibinfo {author} {\bibfnamefont {J.}~\bibnamefont {Chao}},\ and\
  \bibinfo {author} {\bibfnamefont {M.}~\bibnamefont {Huang}},\ }\bibfield
  {title} {\bibinfo {title} {{Effect of the anomalous magnetic moment of quarks
  on magnetized QCD matter and meson spectra}},\ }\href
  {https://doi.org/10.1103/PhysRevD.103.076015} {\bibfield  {journal} {\bibinfo
   {journal} {Phys. Rev. D}\ }\textbf {\bibinfo {volume} {103}},\ \bibinfo
  {pages} {076015} (\bibinfo {year} {2021})},\ \Eprint
  {https://arxiv.org/abs/2007.13122} {arXiv:2007.13122 [hep-ph]} \BibitemShut
  {NoStop}%
\bibitem [{\citenamefont {Lin}\ \emph {et~al.}(2022)\citenamefont {Lin},
  \citenamefont {Xu},\ and\ \citenamefont {Huang}}]{Lin:2022ied}%
  \BibitemOpen
  \bibfield  {author} {\bibinfo {author} {\bibfnamefont {F.}~\bibnamefont
  {Lin}}, \bibinfo {author} {\bibfnamefont {K.}~\bibnamefont {Xu}},\ and\
  \bibinfo {author} {\bibfnamefont {M.}~\bibnamefont {Huang}},\ }\bibfield
  {title} {\bibinfo {title} {{Magnetism of QCD matter and the pion mass from
  tensor-type spin polarization and the anomalous magnetic moment of quarks}},\
  }\href {https://doi.org/10.1103/PhysRevD.106.016005} {\bibfield  {journal}
  {\bibinfo  {journal} {Phys. Rev. D}\ }\textbf {\bibinfo {volume} {106}},\
  \bibinfo {pages} {016005} (\bibinfo {year} {2022})},\ \Eprint
  {https://arxiv.org/abs/2202.03226} {arXiv:2202.03226 [hep-ph]} \BibitemShut
  {NoStop}%
\bibitem [{\citenamefont {Kawaguchi}\ and\ \citenamefont
  {Huang}(2023)}]{Kawaguchi:2022dbq}%
  \BibitemOpen
  \bibfield  {author} {\bibinfo {author} {\bibfnamefont {M.}~\bibnamefont
  {Kawaguchi}}\ and\ \bibinfo {author} {\bibfnamefont {M.}~\bibnamefont
  {Huang}},\ }\bibfield  {title} {\bibinfo {title} {{Restriction on the form of
  the quark anomalous magnetic moment from lattice QCD results}},\ }\href
  {https://doi.org/10.1088/1674-1137/acc641} {\bibfield  {journal} {\bibinfo
  {journal} {Chin. Phys. C}\ }\textbf {\bibinfo {volume} {47}},\ \bibinfo
  {pages} {064103} (\bibinfo {year} {2023})},\ \Eprint
  {https://arxiv.org/abs/2205.08169} {arXiv:2205.08169 [hep-ph]} \BibitemShut
  {NoStop}%
\bibitem [{\citenamefont {Tavares}\ \emph {et~al.}(2024)\citenamefont
  {Tavares}, \citenamefont {Avancini}, \citenamefont {Farias},\ and\
  \citenamefont {Cardoso}}]{Tavares:2023oln}%
  \BibitemOpen
  \bibfield  {author} {\bibinfo {author} {\bibfnamefont {W.~R.}\ \bibnamefont
  {Tavares}}, \bibinfo {author} {\bibfnamefont {S.~S.}\ \bibnamefont
  {Avancini}}, \bibinfo {author} {\bibfnamefont {R.~L.~S.}\ \bibnamefont
  {Farias}},\ and\ \bibinfo {author} {\bibfnamefont {R.~P.}\ \bibnamefont
  {Cardoso}},\ }\bibfield  {title} {\bibinfo {title} {{Artificial first-order
  phase transition in a magnetized Nambu\textendash{}Jona-Lasinio model with a
  quark anomalous magnetic moment}},\ }\href
  {https://doi.org/10.1103/PhysRevD.109.016011} {\bibfield  {journal} {\bibinfo
   {journal} {Phys. Rev. D}\ }\textbf {\bibinfo {volume} {109}},\ \bibinfo
  {pages} {016011} (\bibinfo {year} {2024})},\ \Eprint
  {https://arxiv.org/abs/2309.04055} {arXiv:2309.04055 [hep-ph]} \BibitemShut
  {NoStop}%
\bibitem [{\citenamefont {Chaudhuri}\ \emph {et~al.}(2023)\citenamefont
  {Chaudhuri}, \citenamefont {Ghosh}, \citenamefont {Roy},\ and\ \citenamefont
  {Sarkar}}]{Chaudhuri:2023djv}%
  \BibitemOpen
  \bibfield  {author} {\bibinfo {author} {\bibfnamefont {N.}~\bibnamefont
  {Chaudhuri}}, \bibinfo {author} {\bibfnamefont {S.}~\bibnamefont {Ghosh}},
  \bibinfo {author} {\bibfnamefont {P.}~\bibnamefont {Roy}},\ and\ \bibinfo
  {author} {\bibfnamefont {S.}~\bibnamefont {Sarkar}},\ }\bibfield  {title}
  {\bibinfo {title} {{Collective modes of a massive fermion in a magnetized
  medium with finite anomalous magnetic moment}},\ }\href
  {https://doi.org/10.1103/PhysRevD.108.116006} {\bibfield  {journal} {\bibinfo
   {journal} {Phys. Rev. D}\ }\textbf {\bibinfo {volume} {108}},\ \bibinfo
  {pages} {116006} (\bibinfo {year} {2023})},\ \Eprint
  {https://arxiv.org/abs/2310.05769} {arXiv:2310.05769 [hep-ph]} \BibitemShut
  {NoStop}%
\bibitem [{\citenamefont {Islam}\ \emph {et~al.}(2023)\citenamefont {Islam},
  \citenamefont {Ali},\ and\ \citenamefont {Huang}}]{Islam:2023zyo}%
  \BibitemOpen
  \bibfield  {author} {\bibinfo {author} {\bibfnamefont {C.~A.}\ \bibnamefont
  {Islam}}, \bibinfo {author} {\bibfnamefont {M.~S.}\ \bibnamefont {Ali}},\
  and\ \bibinfo {author} {\bibfnamefont {M.}~\bibnamefont {Huang}},\ }\bibfield
   {title} {\bibinfo {title} {{Deciding on the anomalous magnetic moment of
  quarks in a framework of nonlocal NJL model}},\ }\href@noop {} {\bibfield
  {journal} {\bibinfo  {journal} {arXiv preprint arXiv:2302.00696}\ } (\bibinfo
  {year} {2023})},\ \Eprint {https://arxiv.org/abs/2302.00696}
  {arXiv:2302.00696 [hep-ph]} \BibitemShut {NoStop}%
\bibitem [{\citenamefont {Ferreira}\ \emph {et~al.}(2014)\citenamefont
  {Ferreira}, \citenamefont {Costa}, \citenamefont {Louren\c{c}o},
  \citenamefont {Frederico},\ and\ \citenamefont
  {Provid\^encia}}]{Ferreira:2014kpa}%
  \BibitemOpen
  \bibfield  {author} {\bibinfo {author} {\bibfnamefont {M.}~\bibnamefont
  {Ferreira}}, \bibinfo {author} {\bibfnamefont {P.}~\bibnamefont {Costa}},
  \bibinfo {author} {\bibfnamefont {O.}~\bibnamefont {Louren\c{c}o}}, \bibinfo
  {author} {\bibfnamefont {T.}~\bibnamefont {Frederico}},\ and\ \bibinfo
  {author} {\bibfnamefont {C.}~\bibnamefont {Provid\^encia}},\ }\bibfield
  {title} {\bibinfo {title} {{Inverse magnetic catalysis in the (2+1)-flavor
  Nambu-Jona-Lasinio and Polyakov-Nambu-Jona-Lasinio models}},\ }\href
  {https://doi.org/10.1103/PhysRevD.89.116011} {\bibfield  {journal} {\bibinfo
  {journal} {Phys. Rev. D}\ }\textbf {\bibinfo {volume} {89}},\ \bibinfo
  {pages} {116011} (\bibinfo {year} {2014})},\ \Eprint
  {https://arxiv.org/abs/1404.5577} {arXiv:1404.5577 [hep-ph]} \BibitemShut
  {NoStop}%
\bibitem [{\citenamefont {Liu}\ \emph {et~al.}(2016)\citenamefont {Liu},
  \citenamefont {Yu}, \citenamefont {Chernodub},\ and\ \citenamefont
  {Huang}}]{Liu:2016vuw}%
  \BibitemOpen
  \bibfield  {author} {\bibinfo {author} {\bibfnamefont {H.}~\bibnamefont
  {Liu}}, \bibinfo {author} {\bibfnamefont {L.}~\bibnamefont {Yu}}, \bibinfo
  {author} {\bibfnamefont {M.}~\bibnamefont {Chernodub}},\ and\ \bibinfo
  {author} {\bibfnamefont {M.}~\bibnamefont {Huang}},\ }\bibfield  {title}
  {\bibinfo {title} {{Possible formation of high temperature superconductor at
  an early stage of heavy-ion collisions}},\ }\href
  {https://doi.org/10.1103/PhysRevD.94.113006} {\bibfield  {journal} {\bibinfo
  {journal} {Phys. Rev. D}\ }\textbf {\bibinfo {volume} {94}},\ \bibinfo
  {pages} {113006} (\bibinfo {year} {2016})},\ \Eprint
  {https://arxiv.org/abs/1604.06662} {arXiv:1604.06662 [hep-ph]} \BibitemShut
  {NoStop}%
\bibitem [{\citenamefont {Avancini}\ \emph {et~al.}(2017)\citenamefont
  {Avancini}, \citenamefont {Farias}, \citenamefont {Benghi~Pinto},
  \citenamefont {Tavares},\ and\ \citenamefont {Tim\'oteo}}]{Avancini:2016fgq}%
  \BibitemOpen
  \bibfield  {author} {\bibinfo {author} {\bibfnamefont {S.~S.}\ \bibnamefont
  {Avancini}}, \bibinfo {author} {\bibfnamefont {R.~L.~S.}\ \bibnamefont
  {Farias}}, \bibinfo {author} {\bibfnamefont {M.}~\bibnamefont
  {Benghi~Pinto}}, \bibinfo {author} {\bibfnamefont {W.~R.}\ \bibnamefont
  {Tavares}},\ and\ \bibinfo {author} {\bibfnamefont {V.~S.}\ \bibnamefont
  {Tim\'oteo}},\ }\bibfield  {title} {\bibinfo {title} {{$\pi_0$ pole mass
  calculation in a strong magnetic field and lattice constraints}},\ }\href
  {https://doi.org/10.1016/j.physletb.2017.02.002} {\bibfield  {journal}
  {\bibinfo  {journal} {Phys. Lett. B}\ }\textbf {\bibinfo {volume} {767}},\
  \bibinfo {pages} {247} (\bibinfo {year} {2017})},\ \Eprint
  {https://arxiv.org/abs/1606.05754} {arXiv:1606.05754 [hep-ph]} \BibitemShut
  {NoStop}%
\bibitem [{\citenamefont {Ayala}\ \emph {et~al.}(2016)\citenamefont {Ayala},
  \citenamefont {Dominguez}, \citenamefont {Hernandez}, \citenamefont {Loewe},
  \citenamefont {Raya}, \citenamefont {Rojas},\ and\ \citenamefont
  {Villavicencio}}]{Ayala:2016bbi}%
  \BibitemOpen
  \bibfield  {author} {\bibinfo {author} {\bibfnamefont {A.}~\bibnamefont
  {Ayala}}, \bibinfo {author} {\bibfnamefont {C.~A.}\ \bibnamefont
  {Dominguez}}, \bibinfo {author} {\bibfnamefont {L.~A.}\ \bibnamefont
  {Hernandez}}, \bibinfo {author} {\bibfnamefont {M.}~\bibnamefont {Loewe}},
  \bibinfo {author} {\bibfnamefont {A.}~\bibnamefont {Raya}}, \bibinfo {author}
  {\bibfnamefont {J.~C.}\ \bibnamefont {Rojas}},\ and\ \bibinfo {author}
  {\bibfnamefont {C.}~\bibnamefont {Villavicencio}},\ }\bibfield  {title}
  {\bibinfo {title} {{Thermomagnetic properties of the strong coupling in the
  local Nambu\textendash{}Jona-Lasinio model}},\ }\href
  {https://doi.org/10.1103/PhysRevD.94.054019} {\bibfield  {journal} {\bibinfo
  {journal} {Phys. Rev. D}\ }\textbf {\bibinfo {volume} {94}},\ \bibinfo
  {pages} {054019} (\bibinfo {year} {2016})},\ \Eprint
  {https://arxiv.org/abs/1603.00833} {arXiv:1603.00833 [hep-ph]} \BibitemShut
  {NoStop}%
\bibitem [{\citenamefont {Farias}\ \emph {et~al.}(2017)\citenamefont {Farias},
  \citenamefont {Timoteo}, \citenamefont {Avancini}, \citenamefont {Pinto},\
  and\ \citenamefont {Krein}}]{Farias:2016gmy}%
  \BibitemOpen
  \bibfield  {author} {\bibinfo {author} {\bibfnamefont {R.~L.~S.}\
  \bibnamefont {Farias}}, \bibinfo {author} {\bibfnamefont {V.~S.}\
  \bibnamefont {Timoteo}}, \bibinfo {author} {\bibfnamefont {S.~S.}\
  \bibnamefont {Avancini}}, \bibinfo {author} {\bibfnamefont {M.~B.}\
  \bibnamefont {Pinto}},\ and\ \bibinfo {author} {\bibfnamefont
  {G.}~\bibnamefont {Krein}},\ }\bibfield  {title} {\bibinfo {title}
  {{Thermo-magnetic effects in quark matter: Nambu--Jona-Lasinio model
  constrained by lattice QCD}},\ }\href
  {https://doi.org/10.1140/epja/i2017-12320-8} {\bibfield  {journal} {\bibinfo
  {journal} {Eur. Phys. J. A}\ }\textbf {\bibinfo {volume} {53}},\ \bibinfo
  {pages} {101} (\bibinfo {year} {2017})},\ \Eprint
  {https://arxiv.org/abs/1603.03847} {arXiv:1603.03847 [hep-ph]} \BibitemShut
  {NoStop}%
\bibitem [{\citenamefont {Chen}\ \emph {et~al.}(2022)\citenamefont {Chen},
  \citenamefont {Zhang},\ and\ \citenamefont {Hou}}]{Chen:2021gop}%
  \BibitemOpen
  \bibfield  {author} {\bibinfo {author} {\bibfnamefont {X.}~\bibnamefont
  {Chen}}, \bibinfo {author} {\bibfnamefont {L.}~\bibnamefont {Zhang}},\ and\
  \bibinfo {author} {\bibfnamefont {D.}~\bibnamefont {Hou}},\ }\bibfield
  {title} {\bibinfo {title} {{Running coupling constant at finite chemical
  potential and magnetic field from holography}},\ }\href
  {https://doi.org/10.1088/1674-1137/ac5c2d} {\bibfield  {journal} {\bibinfo
  {journal} {Chin. Phys. C}\ }\textbf {\bibinfo {volume} {46}},\ \bibinfo
  {pages} {073101} (\bibinfo {year} {2022})},\ \Eprint
  {https://arxiv.org/abs/2108.03840} {arXiv:2108.03840 [hep-ph]} \BibitemShut
  {NoStop}%
\bibitem [{\citenamefont {Sheng}\ \emph {et~al.}(2022)\citenamefont {Sheng},
  \citenamefont {Wang},\ and\ \citenamefont {Yu}}]{Sheng:2021evj}%
  \BibitemOpen
  \bibfield  {author} {\bibinfo {author} {\bibfnamefont {B.-k.}\ \bibnamefont
  {Sheng}}, \bibinfo {author} {\bibfnamefont {X.}~\bibnamefont {Wang}},\ and\
  \bibinfo {author} {\bibfnamefont {L.}~\bibnamefont {Yu}},\ }\bibfield
  {title} {\bibinfo {title} {{Impacts of inverse magnetic catalysis on
  screening masses of neutral pions and sigma mesons in hot and magnetized
  quark matter}},\ }\href {https://doi.org/10.1103/PhysRevD.105.034003}
  {\bibfield  {journal} {\bibinfo  {journal} {Phys. Rev. D}\ }\textbf {\bibinfo
  {volume} {105}},\ \bibinfo {pages} {034003} (\bibinfo {year} {2022})},\
  \Eprint {https://arxiv.org/abs/2110.12811} {arXiv:2110.12811 [hep-ph]}
  \BibitemShut {NoStop}%
\bibitem [{\citenamefont {Mao}\ and\ \citenamefont {Tian}(2022)}]{Mao:2022nfs}%
  \BibitemOpen
  \bibfield  {author} {\bibinfo {author} {\bibfnamefont {S.}~\bibnamefont
  {Mao}}\ and\ \bibinfo {author} {\bibfnamefont {Y.}~\bibnamefont {Tian}},\
  }\bibfield  {title} {\bibinfo {title} {{Pion superfluid phase transition
  under an external magnetic field including the inverse magnetic catalysis
  effect}},\ }\href {https://doi.org/10.1103/PhysRevD.106.094017} {\bibfield
  {journal} {\bibinfo  {journal} {Phys. Rev. D}\ }\textbf {\bibinfo {volume}
  {106}},\ \bibinfo {pages} {094017} (\bibinfo {year} {2022})},\ \Eprint
  {https://arxiv.org/abs/2209.10738} {arXiv:2209.10738 [nucl-th]} \BibitemShut
  {NoStop}%
\bibitem [{\citenamefont {Wen}\ \emph {et~al.}(2023)\citenamefont {Wen},
  \citenamefont {Yin}, \citenamefont {Fu},\ and\ \citenamefont
  {Huang}}]{Wen:2023qcz}%
  \BibitemOpen
  \bibfield  {author} {\bibinfo {author} {\bibfnamefont {R.}~\bibnamefont
  {Wen}}, \bibinfo {author} {\bibfnamefont {S.}~\bibnamefont {Yin}}, \bibinfo
  {author} {\bibfnamefont {W.-j.}\ \bibnamefont {Fu}},\ and\ \bibinfo {author}
  {\bibfnamefont {M.}~\bibnamefont {Huang}},\ }\bibfield  {title} {\bibinfo
  {title} {{Functional renormalization group study of neutral and charged pions
  in magnetic fields in the quark-meson model}},\ }\href
  {https://doi.org/10.1103/PhysRevD.108.076020} {\bibfield  {journal} {\bibinfo
   {journal} {Phys. Rev. D}\ }\textbf {\bibinfo {volume} {108}},\ \bibinfo
  {pages} {076020} (\bibinfo {year} {2023})},\ \Eprint
  {https://arxiv.org/abs/2306.04045} {arXiv:2306.04045 [hep-ph]} \BibitemShut
  {NoStop}%
\bibitem [{\citenamefont {Li}\ \emph {et~al.}(2019)\citenamefont {Li},
  \citenamefont {Fu},\ and\ \citenamefont {Liu}}]{Li:2019nzj}%
  \BibitemOpen
  \bibfield  {author} {\bibinfo {author} {\bibfnamefont {X.}~\bibnamefont
  {Li}}, \bibinfo {author} {\bibfnamefont {W.-j.}\ \bibnamefont {Fu}},\ and\
  \bibinfo {author} {\bibfnamefont {Y.-x.}\ \bibnamefont {Liu}},\ }\bibfield
  {title} {\bibinfo {title} {{Thermodynamics of 2+1 Flavor Polyakov-Loop
  Quark-Meson Model under External Magnetic Field}},\ }\href
  {https://doi.org/10.1103/PhysRevD.99.074029} {\bibfield  {journal} {\bibinfo
  {journal} {Phys. Rev. D}\ }\textbf {\bibinfo {volume} {99}},\ \bibinfo
  {pages} {074029} (\bibinfo {year} {2019})},\ \Eprint
  {https://arxiv.org/abs/1902.03866} {arXiv:1902.03866 [hep-ph]} \BibitemShut
  {NoStop}%
\bibitem [{\citenamefont {Kamikado}\ and\ \citenamefont
  {Kanazawa}(2014)}]{Kamikado:2013pya}%
  \BibitemOpen
  \bibfield  {author} {\bibinfo {author} {\bibfnamefont {K.}~\bibnamefont
  {Kamikado}}\ and\ \bibinfo {author} {\bibfnamefont {T.}~\bibnamefont
  {Kanazawa}},\ }\bibfield  {title} {\bibinfo {title} {{Chiral dynamics in a
  magnetic field from the functional renormalization group}},\ }\href
  {https://doi.org/10.1007/JHEP03(2014)009} {\bibfield  {journal} {\bibinfo
  {journal} {JHEP}\ }\textbf {\bibinfo {volume} {03}},\ \bibinfo {pages}
  {009}},\ \Eprint {https://arxiv.org/abs/1312.3124} {arXiv:1312.3124 [hep-ph]}
  \BibitemShut {NoStop}%
\bibitem [{\citenamefont {Kamikado}\ and\ \citenamefont
  {Kanazawa}(2015)}]{Kamikado:2014bua}%
  \BibitemOpen
  \bibfield  {author} {\bibinfo {author} {\bibfnamefont {K.}~\bibnamefont
  {Kamikado}}\ and\ \bibinfo {author} {\bibfnamefont {T.}~\bibnamefont
  {Kanazawa}},\ }\bibfield  {title} {\bibinfo {title} {{Magnetic susceptibility
  of a strongly interacting thermal medium with 2$+$1 quark flavors}},\ }\href
  {https://doi.org/10.1007/JHEP01(2015)129} {\bibfield  {journal} {\bibinfo
  {journal} {JHEP}\ }\textbf {\bibinfo {volume} {01}},\ \bibinfo {pages}
  {129}},\ \Eprint {https://arxiv.org/abs/1410.6253} {arXiv:1410.6253 [hep-ph]}
  \BibitemShut {NoStop}%
\bibitem [{\citenamefont {Fukushima}\ and\ \citenamefont
  {Pawlowski}(2012)}]{Fukushima:2012xw}%
  \BibitemOpen
  \bibfield  {author} {\bibinfo {author} {\bibfnamefont {K.}~\bibnamefont
  {Fukushima}}\ and\ \bibinfo {author} {\bibfnamefont {J.~M.}\ \bibnamefont
  {Pawlowski}},\ }\bibfield  {title} {\bibinfo {title} {{Magnetic catalysis in
  hot and dense quark matter and quantum fluctuations}},\ }\href
  {https://doi.org/10.1103/PhysRevD.86.076013} {\bibfield  {journal} {\bibinfo
  {journal} {Phys. Rev. D}\ }\textbf {\bibinfo {volume} {86}},\ \bibinfo
  {pages} {076013} (\bibinfo {year} {2012})},\ \Eprint
  {https://arxiv.org/abs/1203.4330} {arXiv:1203.4330 [hep-ph]} \BibitemShut
  {NoStop}%
\bibitem [{\citenamefont {Fu}\ and\ \citenamefont {Liu}(2017)}]{Fu:2017vvg}%
  \BibitemOpen
  \bibfield  {author} {\bibinfo {author} {\bibfnamefont {W.-j.}\ \bibnamefont
  {Fu}}\ and\ \bibinfo {author} {\bibfnamefont {Y.-x.}\ \bibnamefont {Liu}},\
  }\bibfield  {title} {\bibinfo {title} {{Four-fermion interactions and the
  chiral symmetry breaking in an external magnetic field}},\ }\href
  {https://doi.org/10.1103/PhysRevD.96.074019} {\bibfield  {journal} {\bibinfo
  {journal} {Phys. Rev. D}\ }\textbf {\bibinfo {volume} {96}},\ \bibinfo
  {pages} {074019} (\bibinfo {year} {2017})},\ \Eprint
  {https://arxiv.org/abs/1705.09841} {arXiv:1705.09841 [hep-ph]} \BibitemShut
  {NoStop}%
\bibitem [{\citenamefont {Bohra}\ \emph {et~al.}(2020)\citenamefont {Bohra},
  \citenamefont {Dudal}, \citenamefont {Hajilou},\ and\ \citenamefont
  {Mahapatra}}]{Bohra:2019ebj}%
  \BibitemOpen
  \bibfield  {author} {\bibinfo {author} {\bibfnamefont {H.}~\bibnamefont
  {Bohra}}, \bibinfo {author} {\bibfnamefont {D.}~\bibnamefont {Dudal}},
  \bibinfo {author} {\bibfnamefont {A.}~\bibnamefont {Hajilou}},\ and\ \bibinfo
  {author} {\bibfnamefont {S.}~\bibnamefont {Mahapatra}},\ }\bibfield  {title}
  {\bibinfo {title} {{Anisotropic string tensions and inversely magnetic
  catalyzed deconfinement from a dynamical AdS/QCD model}},\ }\href
  {https://doi.org/10.1016/j.physletb.2019.135184} {\bibfield  {journal}
  {\bibinfo  {journal} {Phys. Lett. B}\ }\textbf {\bibinfo {volume} {801}},\
  \bibinfo {pages} {135184} (\bibinfo {year} {2020})},\ \Eprint
  {https://arxiv.org/abs/1907.01852} {arXiv:1907.01852 [hep-th]} \BibitemShut
  {NoStop}%
\bibitem [{\citenamefont {Bohra}\ \emph {et~al.}(2021)\citenamefont {Bohra},
  \citenamefont {Dudal}, \citenamefont {Hajilou},\ and\ \citenamefont
  {Mahapatra}}]{Bohra:2020qom}%
  \BibitemOpen
  \bibfield  {author} {\bibinfo {author} {\bibfnamefont {H.}~\bibnamefont
  {Bohra}}, \bibinfo {author} {\bibfnamefont {D.}~\bibnamefont {Dudal}},
  \bibinfo {author} {\bibfnamefont {A.}~\bibnamefont {Hajilou}},\ and\ \bibinfo
  {author} {\bibfnamefont {S.}~\bibnamefont {Mahapatra}},\ }\bibfield  {title}
  {\bibinfo {title} {{Chiral transition in the probe approximation from an
  Einstein-Maxwell-dilaton gravity model}},\ }\href
  {https://doi.org/10.1103/PhysRevD.103.086021} {\bibfield  {journal} {\bibinfo
   {journal} {Phys. Rev. D}\ }\textbf {\bibinfo {volume} {103}},\ \bibinfo
  {pages} {086021} (\bibinfo {year} {2021})},\ \Eprint
  {https://arxiv.org/abs/2010.04578} {arXiv:2010.04578 [hep-th]} \BibitemShut
  {NoStop}%
\bibitem [{\citenamefont {Wen}\ \emph {et~al.}(2024)\citenamefont {Wen},
  \citenamefont {Cao}, \citenamefont {Chao},\ and\ \citenamefont
  {Liu}}]{Wen:2024hgu}%
  \BibitemOpen
  \bibfield  {author} {\bibinfo {author} {\bibfnamefont {N.}~\bibnamefont
  {Wen}}, \bibinfo {author} {\bibfnamefont {X.}~\bibnamefont {Cao}}, \bibinfo
  {author} {\bibfnamefont {J.}~\bibnamefont {Chao}},\ and\ \bibinfo {author}
  {\bibfnamefont {H.}~\bibnamefont {Liu}},\ }\bibfield  {title} {\bibinfo
  {title} {{Neutral pion masses within a hot and magnetized medium in a
  lattice-improved soft-wall AdS/QCD model}},\ }\href@noop {} {\bibfield
  {journal} {\bibinfo  {journal} {arXiv preprint arXiv:2402.06239}\ } (\bibinfo
  {year} {2024})},\ \Eprint {https://arxiv.org/abs/2402.06239}
  {arXiv:2402.06239 [hep-th]} \BibitemShut {NoStop}%
\bibitem [{\citenamefont {Wen}\ \emph {et~al.}(2025)\citenamefont {Wen},
  \citenamefont {Huang},\ and\ \citenamefont {Huang}}]{Wen:2025cpq}%
  \BibitemOpen
  \bibfield  {author} {\bibinfo {author} {\bibfnamefont {R.}~\bibnamefont
  {Wen}}, \bibinfo {author} {\bibfnamefont {C.}~\bibnamefont {Huang}},\ and\
  \bibinfo {author} {\bibfnamefont {M.}~\bibnamefont {Huang}},\ }\bibfield
  {title} {\bibinfo {title} {{Functional renormalization group study of
  anomalous magnetic moment in a low energy effective theory}},\ }\href@noop {}
  {\bibfield  {journal} {\bibinfo  {journal} {arXiv preprint arXiv:2506.20246}\
  } (\bibinfo {year} {2025})},\ \Eprint {https://arxiv.org/abs/2506.20246}
  {arXiv:2506.20246 [hep-ph]} \BibitemShut {NoStop}%
\bibitem [{\citenamefont {Hatsuda}\ and\ \citenamefont
  {Kunihiro}(1994)}]{Hatsuda:1994pi}%
  \BibitemOpen
  \bibfield  {author} {\bibinfo {author} {\bibfnamefont {T.}~\bibnamefont
  {Hatsuda}}\ and\ \bibinfo {author} {\bibfnamefont {T.}~\bibnamefont
  {Kunihiro}},\ }\bibfield  {title} {\bibinfo {title} {{QCD phenomenology based
  on a chiral effective Lagrangian}},\ }\href
  {https://doi.org/10.1016/0370-1573(94)90022-1} {\bibfield  {journal}
  {\bibinfo  {journal} {Phys. Rept.}\ }\textbf {\bibinfo {volume} {247}},\
  \bibinfo {pages} {221} (\bibinfo {year} {1994})},\ \Eprint
  {https://arxiv.org/abs/hep-ph/9401310} {arXiv:hep-ph/9401310} \BibitemShut
  {NoStop}%
\bibitem [{\citenamefont {Stephanov}(2004)}]{Stephanov:2004wx}%
  \BibitemOpen
  \bibfield  {author} {\bibinfo {author} {\bibfnamefont {M.~A.}\ \bibnamefont
  {Stephanov}},\ }\bibfield  {title} {\bibinfo {title} {{QCD Phase Diagram and
  the Critical Point}},\ }\href {https://doi.org/10.1143/PTPS.153.139}
  {\bibfield  {journal} {\bibinfo  {journal} {Prog. Theor. Phys. Suppl.}\
  }\textbf {\bibinfo {volume} {153}},\ \bibinfo {pages} {139} (\bibinfo {year}
  {2004})},\ \Eprint {https://arxiv.org/abs/hep-ph/0402115}
  {arXiv:hep-ph/0402115} \BibitemShut {NoStop}%
\bibitem [{\citenamefont {Gatto}\ and\ \citenamefont
  {Ruggieri}(2011)}]{Gatto:2010pt}%
  \BibitemOpen
  \bibfield  {author} {\bibinfo {author} {\bibfnamefont {R.}~\bibnamefont
  {Gatto}}\ and\ \bibinfo {author} {\bibfnamefont {M.}~\bibnamefont
  {Ruggieri}},\ }\bibfield  {title} {\bibinfo {title} {{Deconfinement and
  Chiral Symmetry Restoration in a Strong Magnetic Background}},\ }\href
  {https://doi.org/10.1103/PhysRevD.83.034016} {\bibfield  {journal} {\bibinfo
  {journal} {Phys. Rev. D}\ }\textbf {\bibinfo {volume} {83}},\ \bibinfo
  {pages} {034016} (\bibinfo {year} {2011})},\ \Eprint
  {https://arxiv.org/abs/1012.1291} {arXiv:1012.1291 [hep-ph]} \BibitemShut
  {NoStop}%
\bibitem [{\citenamefont {Sheng}\ \emph {et~al.}(2021)\citenamefont {Sheng},
  \citenamefont {Wang}, \citenamefont {Wang},\ and\ \citenamefont
  {Yu}}]{Sheng:2020hge}%
  \BibitemOpen
  \bibfield  {author} {\bibinfo {author} {\bibfnamefont {B.}~\bibnamefont
  {Sheng}}, \bibinfo {author} {\bibfnamefont {Y.}~\bibnamefont {Wang}},
  \bibinfo {author} {\bibfnamefont {X.}~\bibnamefont {Wang}},\ and\ \bibinfo
  {author} {\bibfnamefont {L.}~\bibnamefont {Yu}},\ }\bibfield  {title}
  {\bibinfo {title} {{Pole and screening masses of neutral pions in a hot and
  magnetized medium: A comprehensive study in the
  Nambu\textendash{}Jona-Lasinio model}},\ }\href
  {https://doi.org/10.1103/PhysRevD.103.094001} {\bibfield  {journal} {\bibinfo
   {journal} {Phys. Rev. D}\ }\textbf {\bibinfo {volume} {103}},\ \bibinfo
  {pages} {094001} (\bibinfo {year} {2021})},\ \Eprint
  {https://arxiv.org/abs/2010.05716} {arXiv:2010.05716 [hep-ph]} \BibitemShut
  {NoStop}%
\bibitem [{\citenamefont {Mei}\ \emph {et~al.}(2024)\citenamefont {Mei},
  \citenamefont {Wen}, \citenamefont {Mao}, \citenamefont {Huang},\ and\
  \citenamefont {Xu}}]{Mei:2024rjg}%
  \BibitemOpen
  \bibfield  {author} {\bibinfo {author} {\bibfnamefont {J.}~\bibnamefont
  {Mei}}, \bibinfo {author} {\bibfnamefont {R.}~\bibnamefont {Wen}}, \bibinfo
  {author} {\bibfnamefont {S.}~\bibnamefont {Mao}}, \bibinfo {author}
  {\bibfnamefont {M.}~\bibnamefont {Huang}},\ and\ \bibinfo {author}
  {\bibfnamefont {K.}~\bibnamefont {Xu}},\ }\bibfield  {title} {\bibinfo
  {title} {{Magnetic catalysis and diamagnetism from pion fluctuations}},\
  }\href {https://doi.org/10.1103/PhysRevD.110.034024} {\bibfield  {journal}
  {\bibinfo  {journal} {Phys. Rev. D}\ }\textbf {\bibinfo {volume} {110}},\
  \bibinfo {pages} {034024} (\bibinfo {year} {2024})},\ \Eprint
  {https://arxiv.org/abs/2402.19193} {arXiv:2402.19193 [hep-ph]} \BibitemShut
  {NoStop}%
\bibitem [{\citenamefont {Bazavov}\ \emph {et~al.}(2019)\citenamefont {Bazavov}
  \emph {et~al.}}]{Bazavov:2019www}%
  \BibitemOpen
  \bibfield  {author} {\bibinfo {author} {\bibfnamefont {A.}~\bibnamefont
  {Bazavov}} \emph {et~al.},\ }\bibfield  {title} {\bibinfo {title} {{Meson
  screening masses in (2+1)-flavor QCD}},\ }\href
  {https://doi.org/10.1103/PhysRevD.100.094510} {\bibfield  {journal} {\bibinfo
   {journal} {Phys. Rev. D}\ }\textbf {\bibinfo {volume} {100}},\ \bibinfo
  {pages} {094510} (\bibinfo {year} {2019})},\ \Eprint
  {https://arxiv.org/abs/1908.09552} {arXiv:1908.09552 [hep-lat]} \BibitemShut
  {NoStop}%
\bibitem [{\citenamefont {Ding}\ \emph {et~al.}(2021)\citenamefont {Ding},
  \citenamefont {Li}, \citenamefont {Tomiya}, \citenamefont {Wang},\ and\
  \citenamefont {Zhang}}]{Ding:2020hxw}%
  \BibitemOpen
  \bibfield  {author} {\bibinfo {author} {\bibfnamefont {H.~T.}\ \bibnamefont
  {Ding}}, \bibinfo {author} {\bibfnamefont {S.~T.}\ \bibnamefont {Li}},
  \bibinfo {author} {\bibfnamefont {A.}~\bibnamefont {Tomiya}}, \bibinfo
  {author} {\bibfnamefont {X.~D.}\ \bibnamefont {Wang}},\ and\ \bibinfo
  {author} {\bibfnamefont {Y.}~\bibnamefont {Zhang}},\ }\bibfield  {title}
  {\bibinfo {title} {{Chiral properties of (2+1)-flavor QCD in strong magnetic
  fields at zero temperature}},\ }\href
  {https://doi.org/10.1103/PhysRevD.104.014505} {\bibfield  {journal} {\bibinfo
   {journal} {Phys. Rev. D}\ }\textbf {\bibinfo {volume} {104}},\ \bibinfo
  {pages} {014505} (\bibinfo {year} {2021})},\ \Eprint
  {https://arxiv.org/abs/2008.00493} {arXiv:2008.00493 [hep-lat]} \BibitemShut
  {NoStop}%
\bibitem [{\citenamefont {Laudicina}\ \emph {et~al.}(2022)\citenamefont
  {Laudicina}, \citenamefont {Dalla~Brida}, \citenamefont {Giusti},
  \citenamefont {Harris},\ and\ \citenamefont {Pepe}}]{Laudicina:2021gex}%
  \BibitemOpen
  \bibfield  {author} {\bibinfo {author} {\bibfnamefont {D.}~\bibnamefont
  {Laudicina}}, \bibinfo {author} {\bibfnamefont {M.}~\bibnamefont
  {Dalla~Brida}}, \bibinfo {author} {\bibfnamefont {L.}~\bibnamefont {Giusti}},
  \bibinfo {author} {\bibfnamefont {T.}~\bibnamefont {Harris}},\ and\ \bibinfo
  {author} {\bibfnamefont {M.}~\bibnamefont {Pepe}},\ }\bibfield  {title}
  {\bibinfo {title} {{Computation of QCD meson screening masses at high
  temperature}},\ }\href {https://doi.org/10.22323/1.396.0190} {\bibfield
  {journal} {\bibinfo  {journal} {PoS}\ }\textbf {\bibinfo {volume}
  {LATTICE2021}},\ \bibinfo {pages} {190} (\bibinfo {year} {2022})},\ \Eprint
  {https://arxiv.org/abs/2112.06662} {arXiv:2112.06662 [hep-lat]} \BibitemShut
  {NoStop}%
\bibitem [{\citenamefont {Ishii}\ \emph {et~al.}(2014)\citenamefont {Ishii},
  \citenamefont {Sasaki}, \citenamefont {Kashiwa}, \citenamefont {Kouno},\ and\
  \citenamefont {Yahiro}}]{Ishii:2013kaa}%
  \BibitemOpen
  \bibfield  {author} {\bibinfo {author} {\bibfnamefont {M.}~\bibnamefont
  {Ishii}}, \bibinfo {author} {\bibfnamefont {T.}~\bibnamefont {Sasaki}},
  \bibinfo {author} {\bibfnamefont {K.}~\bibnamefont {Kashiwa}}, \bibinfo
  {author} {\bibfnamefont {H.}~\bibnamefont {Kouno}},\ and\ \bibinfo {author}
  {\bibfnamefont {M.}~\bibnamefont {Yahiro}},\ }\bibfield  {title} {\bibinfo
  {title} {{Effective model approach to meson screening masses at finite
  temperature}},\ }\href {https://doi.org/10.1103/PhysRevD.89.071901}
  {\bibfield  {journal} {\bibinfo  {journal} {Phys. Rev. D}\ }\textbf {\bibinfo
  {volume} {89}},\ \bibinfo {pages} {071901} (\bibinfo {year} {2014})},\
  \Eprint {https://arxiv.org/abs/1312.7424} {arXiv:1312.7424 [hep-ph]}
  \BibitemShut {NoStop}%
\bibitem [{\citenamefont {Czerski}\ and\ \citenamefont
  {Alberico}(2013)}]{Czerski:2012fg}%
  \BibitemOpen
  \bibfield  {author} {\bibinfo {author} {\bibfnamefont {P.}~\bibnamefont
  {Czerski}}\ and\ \bibinfo {author} {\bibfnamefont {W.~M.}\ \bibnamefont
  {Alberico}},\ }\bibfield  {title} {\bibinfo {title} {{Screening Masses of
  Scalar and Pseudo-scalar Excitations in Quark-gluon Plasma}},\ }\href
  {https://doi.org/10.1016/j.nuclphysa.2013.09.008} {\bibfield  {journal}
  {\bibinfo  {journal} {Nucl. Phys. A}\ }\textbf {\bibinfo {volume} {918}},\
  \bibinfo {pages} {170} (\bibinfo {year} {2013})},\ \Eprint
  {https://arxiv.org/abs/1210.6531} {arXiv:1210.6531 [hep-ph]} \BibitemShut
  {NoStop}%
\bibitem [{\citenamefont {Mao}(2016)}]{Mao:2016fha}%
  \BibitemOpen
  \bibfield  {author} {\bibinfo {author} {\bibfnamefont {S.}~\bibnamefont
  {Mao}},\ }\bibfield  {title} {\bibinfo {title} {{Inverse magnetic catalysis
  in Nambu\textendash{}Jona-Lasinio model beyond mean field}},\ }\href
  {https://doi.org/10.1016/j.physletb.2016.05.018} {\bibfield  {journal}
  {\bibinfo  {journal} {Phys. Lett. B}\ }\textbf {\bibinfo {volume} {758}},\
  \bibinfo {pages} {195} (\bibinfo {year} {2016})},\ \Eprint
  {https://arxiv.org/abs/1602.06503} {arXiv:1602.06503 [hep-ph]} \BibitemShut
  {NoStop}%
\bibitem [{\citenamefont {Mao}(2018)}]{Mao:2017tcf}%
  \BibitemOpen
  \bibfield  {author} {\bibinfo {author} {\bibfnamefont {S.}~\bibnamefont
  {Mao}},\ }\bibfield  {title} {\bibinfo {title} {{Chiral Symmetry Restoration
  and Quark Deconfinement beyond Mean Field in a Magnetized PNJL Model}},\
  }\href {https://doi.org/10.1103/PhysRevD.97.011501} {\bibfield  {journal}
  {\bibinfo  {journal} {Phys. Rev. D}\ }\textbf {\bibinfo {volume} {97}},\
  \bibinfo {pages} {011501} (\bibinfo {year} {2018})},\ \Eprint
  {https://arxiv.org/abs/1712.06062} {arXiv:1712.06062 [nucl-th]} \BibitemShut
  {NoStop}%
\bibitem [{\citenamefont {Avancini}\ \emph {et~al.}(2019)\citenamefont
  {Avancini}, \citenamefont {Farias},\ and\ \citenamefont
  {Tavares}}]{Avancini:2018svs}%
  \BibitemOpen
  \bibfield  {author} {\bibinfo {author} {\bibfnamefont {S.~S.}\ \bibnamefont
  {Avancini}}, \bibinfo {author} {\bibfnamefont {R.~L.~S.}\ \bibnamefont
  {Farias}},\ and\ \bibinfo {author} {\bibfnamefont {W.~R.}\ \bibnamefont
  {Tavares}},\ }\bibfield  {title} {\bibinfo {title} {{Neutral meson properties
  in hot and magnetized quark matter: a new magnetic field independent
  regularization scheme applied to NJL-type model}},\ }\href
  {https://doi.org/10.1103/PhysRevD.99.056009} {\bibfield  {journal} {\bibinfo
  {journal} {Phys. Rev. D}\ }\textbf {\bibinfo {volume} {99}},\ \bibinfo
  {pages} {056009} (\bibinfo {year} {2019})},\ \Eprint
  {https://arxiv.org/abs/1812.00945} {arXiv:1812.00945 [hep-ph]} \BibitemShut
  {NoStop}%
\bibitem [{\citenamefont {Mao}(2021)}]{Mao:2019avr}%
  \BibitemOpen
  \bibfield  {author} {\bibinfo {author} {\bibfnamefont {S.}~\bibnamefont
  {Mao}},\ }\bibfield  {title} {\bibinfo {title} {{Chiral crossover
  characterized by Mott transition at finite temperature}},\ }\href
  {https://doi.org/10.1088/1674-1137/abcfad} {\bibfield  {journal} {\bibinfo
  {journal} {Chin. Phys. C}\ }\textbf {\bibinfo {volume} {45}},\ \bibinfo
  {pages} {021004} (\bibinfo {year} {2021})},\ \Eprint
  {https://arxiv.org/abs/1908.02851} {arXiv:1908.02851 [nucl-th]} \BibitemShut
  {NoStop}%
\bibitem [{\citenamefont {Mei}\ \emph {et~al.}(2023)\citenamefont {Mei},
  \citenamefont {Xia},\ and\ \citenamefont {Mao}}]{Mei:2022dkd}%
  \BibitemOpen
  \bibfield  {author} {\bibinfo {author} {\bibfnamefont {J.}~\bibnamefont
  {Mei}}, \bibinfo {author} {\bibfnamefont {T.}~\bibnamefont {Xia}},\ and\
  \bibinfo {author} {\bibfnamefont {S.}~\bibnamefont {Mao}},\ }\bibfield
  {title} {\bibinfo {title} {{Mass spectra of neutral mesons
  K0,\ensuremath{\pi}0,\ensuremath{\eta},\ensuremath{\eta}' at finite magnetic
  field, temperature and quark chemical potential}},\ }\href
  {https://doi.org/10.1103/PhysRevD.107.074018} {\bibfield  {journal} {\bibinfo
   {journal} {Phys. Rev. D}\ }\textbf {\bibinfo {volume} {107}},\ \bibinfo
  {pages} {074018} (\bibinfo {year} {2023})},\ \Eprint
  {https://arxiv.org/abs/2212.04778} {arXiv:2212.04778 [hep-ph]} \BibitemShut
  {NoStop}%
\bibitem [{\citenamefont {Tripolt}\ \emph
  {et~al.}(2014{\natexlab{a}})\citenamefont {Tripolt}, \citenamefont
  {Strodthoff}, \citenamefont {von Smekal},\ and\ \citenamefont
  {Wambach}}]{Tripolt:2013haa}%
  \BibitemOpen
  \bibfield  {author} {\bibinfo {author} {\bibfnamefont {R.-A.}\ \bibnamefont
  {Tripolt}}, \bibinfo {author} {\bibfnamefont {N.}~\bibnamefont {Strodthoff}},
  \bibinfo {author} {\bibfnamefont {L.}~\bibnamefont {von Smekal}},\ and\
  \bibinfo {author} {\bibfnamefont {J.}~\bibnamefont {Wambach}},\ }\bibfield
  {title} {\bibinfo {title} {{Finite-Temperature Spectral Functions from the
  Functional Renormalization Group}},\ }\href
  {https://doi.org/10.22323/1.187.0457} {\bibfield  {journal} {\bibinfo
  {journal} {PoS}\ }\textbf {\bibinfo {volume} {LATTICE2013}},\ \bibinfo
  {pages} {457} (\bibinfo {year} {2014}{\natexlab{a}})},\ \Eprint
  {https://arxiv.org/abs/1311.4304} {arXiv:1311.4304 [hep-lat]} \BibitemShut
  {NoStop}%
\bibitem [{\citenamefont {Tripolt}\ \emph
  {et~al.}(2014{\natexlab{b}})\citenamefont {Tripolt}, \citenamefont {von
  Smekal},\ and\ \citenamefont {Wambach}}]{Tripolt:2014wra}%
  \BibitemOpen
  \bibfield  {author} {\bibinfo {author} {\bibfnamefont {R.-A.}\ \bibnamefont
  {Tripolt}}, \bibinfo {author} {\bibfnamefont {L.}~\bibnamefont {von
  Smekal}},\ and\ \bibinfo {author} {\bibfnamefont {J.}~\bibnamefont
  {Wambach}},\ }\bibfield  {title} {\bibinfo {title} {{Flow equations for
  spectral functions at finite external momenta}},\ }\href
  {https://doi.org/10.1103/PhysRevD.90.074031} {\bibfield  {journal} {\bibinfo
  {journal} {Phys. Rev. D}\ }\textbf {\bibinfo {volume} {90}},\ \bibinfo
  {pages} {074031} (\bibinfo {year} {2014}{\natexlab{b}})},\ \Eprint
  {https://arxiv.org/abs/1408.3512} {arXiv:1408.3512 [hep-ph]} \BibitemShut
  {NoStop}%
\bibitem [{\citenamefont {Wambach}\ \emph {et~al.}(2014)\citenamefont
  {Wambach}, \citenamefont {Tripolt}, \citenamefont {Strodthoff},\ and\
  \citenamefont {von Smekal}}]{Wambach:2014vta}%
  \BibitemOpen
  \bibfield  {author} {\bibinfo {author} {\bibfnamefont {J.}~\bibnamefont
  {Wambach}}, \bibinfo {author} {\bibfnamefont {R.-A.}\ \bibnamefont
  {Tripolt}}, \bibinfo {author} {\bibfnamefont {N.}~\bibnamefont
  {Strodthoff}},\ and\ \bibinfo {author} {\bibfnamefont {L.}~\bibnamefont {von
  Smekal}},\ }\bibfield  {title} {\bibinfo {title} {{Spectral Functions from
  the Functional Renormalization Group}},\ }\href
  {https://doi.org/10.1016/j.nuclphysa.2014.04.027} {\bibfield  {journal}
  {\bibinfo  {journal} {Nucl. Phys. A}\ }\textbf {\bibinfo {volume} {928}},\
  \bibinfo {pages} {156} (\bibinfo {year} {2014})},\ \Eprint
  {https://arxiv.org/abs/1404.7312} {arXiv:1404.7312 [hep-ph]} \BibitemShut
  {NoStop}%
\bibitem [{\citenamefont {Tripolt}\ \emph {et~al.}(2017)\citenamefont
  {Tripolt}, \citenamefont {von Smekal},\ and\ \citenamefont
  {Wambach}}]{Tripolt:2016cey}%
  \BibitemOpen
  \bibfield  {author} {\bibinfo {author} {\bibfnamefont {R.-A.}\ \bibnamefont
  {Tripolt}}, \bibinfo {author} {\bibfnamefont {L.}~\bibnamefont {von
  Smekal}},\ and\ \bibinfo {author} {\bibfnamefont {J.}~\bibnamefont
  {Wambach}},\ }\bibfield  {title} {\bibinfo {title} {{Spectral functions and
  in-medium properties of hadrons}},\ }\href
  {https://doi.org/10.1142/S0218301317400286} {\bibfield  {journal} {\bibinfo
  {journal} {Int. J. Mod. Phys. E}\ }\textbf {\bibinfo {volume} {26}},\
  \bibinfo {pages} {1740028} (\bibinfo {year} {2017})},\ \Eprint
  {https://arxiv.org/abs/1605.00771} {arXiv:1605.00771 [hep-ph]} \BibitemShut
  {NoStop}%
\bibitem [{\citenamefont {Islam}\ \emph {et~al.}(2019)\citenamefont {Islam},
  \citenamefont {Bandyopadhyay}, \citenamefont {Roy},\ and\ \citenamefont
  {Sarkar}}]{Islam:2018sog}%
  \BibitemOpen
  \bibfield  {author} {\bibinfo {author} {\bibfnamefont {C.~A.}\ \bibnamefont
  {Islam}}, \bibinfo {author} {\bibfnamefont {A.}~\bibnamefont
  {Bandyopadhyay}}, \bibinfo {author} {\bibfnamefont {P.~K.}\ \bibnamefont
  {Roy}},\ and\ \bibinfo {author} {\bibfnamefont {S.}~\bibnamefont {Sarkar}},\
  }\bibfield  {title} {\bibinfo {title} {{Spectral function and dilepton rate
  from a strongly magnetized hot and dense medium in light of mean field
  models}},\ }\href {https://doi.org/10.1103/PhysRevD.99.094028} {\bibfield
  {journal} {\bibinfo  {journal} {Phys. Rev. D}\ }\textbf {\bibinfo {volume}
  {99}},\ \bibinfo {pages} {094028} (\bibinfo {year} {2019})},\ \Eprint
  {https://arxiv.org/abs/1812.10380} {arXiv:1812.10380 [hep-ph]} \BibitemShut
  {NoStop}%
\bibitem [{\citenamefont {Wei}\ \emph {et~al.}(2022)\citenamefont {Wei},
  \citenamefont {Islam},\ and\ \citenamefont {Huang}}]{Wei:2021dib}%
  \BibitemOpen
  \bibfield  {author} {\bibinfo {author} {\bibfnamefont {M.}~\bibnamefont
  {Wei}}, \bibinfo {author} {\bibfnamefont {C.~A.}\ \bibnamefont {Islam}},\
  and\ \bibinfo {author} {\bibfnamefont {M.}~\bibnamefont {Huang}},\ }\bibfield
   {title} {\bibinfo {title} {{Production rate and ellipticity of lepton pairs
  from a rotating hot and dense QCD medium}},\ }\href
  {https://doi.org/10.1103/PhysRevD.105.054014} {\bibfield  {journal} {\bibinfo
   {journal} {Phys. Rev. D}\ }\textbf {\bibinfo {volume} {105}},\ \bibinfo
  {pages} {054014} (\bibinfo {year} {2022})},\ \Eprint
  {https://arxiv.org/abs/2111.05192} {arXiv:2111.05192 [hep-ph]} \BibitemShut
  {NoStop}%
\bibitem [{\citenamefont {Wei}\ and\ \citenamefont
  {Huang}(2023)}]{Wei:2023pdf}%
  \BibitemOpen
  \bibfield  {author} {\bibinfo {author} {\bibfnamefont {M.}~\bibnamefont
  {Wei}}\ and\ \bibinfo {author} {\bibfnamefont {M.}~\bibnamefont {Huang}},\
  }\bibfield  {title} {\bibinfo {title} {{Spin alignment of vector mesons from
  quark dynamics in a rotating medium*}},\ }\href
  {https://doi.org/10.1088/1674-1137/acf036} {\bibfield  {journal} {\bibinfo
  {journal} {Chin. Phys. C}\ }\textbf {\bibinfo {volume} {47}},\ \bibinfo
  {pages} {104105} (\bibinfo {year} {2023})},\ \Eprint
  {https://arxiv.org/abs/2303.01897} {arXiv:2303.01897 [hep-ph]} \BibitemShut
  {NoStop}%
\bibitem [{\citenamefont {Ghosh}\ \emph {et~al.}(2024)\citenamefont {Ghosh},
  \citenamefont {Chaudhuri}, \citenamefont {Sarkar},\ and\ \citenamefont
  {Roy}}]{Ghosh:2023rft}%
  \BibitemOpen
  \bibfield  {author} {\bibinfo {author} {\bibfnamefont {S.}~\bibnamefont
  {Ghosh}}, \bibinfo {author} {\bibfnamefont {N.}~\bibnamefont {Chaudhuri}},
  \bibinfo {author} {\bibfnamefont {S.}~\bibnamefont {Sarkar}},\ and\ \bibinfo
  {author} {\bibfnamefont {P.}~\bibnamefont {Roy}},\ }\bibfield  {title}
  {\bibinfo {title} {{Mass and spectral function of scalar and pseudoscalar
  mesons in a hot and chirally imbalanced medium using the two-flavor NJL
  model}},\ }\href {https://doi.org/10.1103/PhysRevD.109.016021} {\bibfield
  {journal} {\bibinfo  {journal} {Phys. Rev. D}\ }\textbf {\bibinfo {volume}
  {109}},\ \bibinfo {pages} {016021} (\bibinfo {year} {2024})},\ \Eprint
  {https://arxiv.org/abs/2312.08652} {arXiv:2312.08652 [hep-ph]} \BibitemShut
  {NoStop}%
\bibitem [{\citenamefont {Bellac}(2011)}]{Bellac:2011kqa}%
  \BibitemOpen
  \bibfield  {author} {\bibinfo {author} {\bibfnamefont {M.~L.}\ \bibnamefont
  {Bellac}},\ }\href {https://doi.org/10.1017/CBO9780511721700} {\emph
  {\bibinfo {title} {{Thermal Field Theory}}}},\ Cambridge Monographs on
  Mathematical Physics\ (\bibinfo  {publisher} {Cambridge University Press},\
  \bibinfo {year} {2011})\BibitemShut {NoStop}%
\bibitem [{\citenamefont {Patkos}\ \emph {et~al.}(2002)\citenamefont {Patkos},
  \citenamefont {Szep},\ and\ \citenamefont {Szepfalusy}}]{Patkos:2002vr}%
  \BibitemOpen
  \bibfield  {author} {\bibinfo {author} {\bibfnamefont {A.}~\bibnamefont
  {Patkos}}, \bibinfo {author} {\bibfnamefont {Z.}~\bibnamefont {Szep}},\ and\
  \bibinfo {author} {\bibfnamefont {P.}~\bibnamefont {Szepfalusy}},\ }\bibfield
   {title} {\bibinfo {title} {{Second sheet sigma pole and the threshold
  enhancement of the spectral function in the scalar isoscalar meson sector}},\
  }\href {https://doi.org/10.1103/PhysRevD.66.116004} {\bibfield  {journal}
  {\bibinfo  {journal} {Phys. Rev. D}\ }\textbf {\bibinfo {volume} {66}},\
  \bibinfo {pages} {116004} (\bibinfo {year} {2002})},\ \Eprint
  {https://arxiv.org/abs/hep-ph/0206040} {arXiv:hep-ph/0206040} \BibitemShut
  {NoStop}%
\bibitem [{\citenamefont {Patkos}\ \emph {et~al.}(2003)\citenamefont {Patkos},
  \citenamefont {Szep},\ and\ \citenamefont {Szepfalusy}}]{Patkos:2003cu}%
  \BibitemOpen
  \bibfield  {author} {\bibinfo {author} {\bibfnamefont {A.}~\bibnamefont
  {Patkos}}, \bibinfo {author} {\bibfnamefont {Z.}~\bibnamefont {Szep}},\ and\
  \bibinfo {author} {\bibfnamefont {P.}~\bibnamefont {Szepfalusy}},\ }\bibfield
   {title} {\bibinfo {title} {{Universal threshold enhancement}},\ }\href
  {https://doi.org/10.1103/PhysRevD.68.047701} {\bibfield  {journal} {\bibinfo
  {journal} {Phys. Rev. D}\ }\textbf {\bibinfo {volume} {68}},\ \bibinfo
  {pages} {047701} (\bibinfo {year} {2003})},\ \Eprint
  {https://arxiv.org/abs/hep-ph/0305100} {arXiv:hep-ph/0305100} \BibitemShut
  {NoStop}%
\bibitem [{\citenamefont {Chiku}\ and\ \citenamefont
  {Hatsuda}(1998)}]{Chiku:1997va}%
  \BibitemOpen
  \bibfield  {author} {\bibinfo {author} {\bibfnamefont {S.}~\bibnamefont
  {Chiku}}\ and\ \bibinfo {author} {\bibfnamefont {T.}~\bibnamefont
  {Hatsuda}},\ }\bibfield  {title} {\bibinfo {title} {{Soft modes associated
  with chiral transition at finite temperature}},\ }\href
  {https://doi.org/10.1103/PhysRevD.57.R6} {\bibfield  {journal} {\bibinfo
  {journal} {Phys. Rev. D}\ }\textbf {\bibinfo {volume} {57}},\ \bibinfo
  {pages} {6} (\bibinfo {year} {1998})},\ \Eprint
  {https://arxiv.org/abs/hep-ph/9706453} {arXiv:hep-ph/9706453} \BibitemShut
  {NoStop}%
\bibitem [{\citenamefont {Sheng}\ \emph
  {et~al.}(2024{\natexlab{a}})\citenamefont {Sheng}, \citenamefont {Yang},
  \citenamefont {Zou},\ and\ \citenamefont {Hou}}]{Sheng:2022ssp}%
  \BibitemOpen
  \bibfield  {author} {\bibinfo {author} {\bibfnamefont {X.-L.}\ \bibnamefont
  {Sheng}}, \bibinfo {author} {\bibfnamefont {S.-Y.}\ \bibnamefont {Yang}},
  \bibinfo {author} {\bibfnamefont {Y.-L.}\ \bibnamefont {Zou}},\ and\ \bibinfo
  {author} {\bibfnamefont {D.}~\bibnamefont {Hou}},\ }\bibfield  {title}
  {\bibinfo {title} {{Mass splitting and spin alignment for $\phi $ mesons in a
  magnetic field in NJL model}},\ }\href
  {https://doi.org/10.1140/epjc/s10052-024-12643-7} {\bibfield  {journal}
  {\bibinfo  {journal} {Eur. Phys. J. C}\ }\textbf {\bibinfo {volume} {84}},\
  \bibinfo {pages} {299} (\bibinfo {year} {2024}{\natexlab{a}})},\ \Eprint
  {https://arxiv.org/abs/2209.01872} {arXiv:2209.01872 [nucl-th]} \BibitemShut
  {NoStop}%
\bibitem [{\citenamefont {Sheng}\ \emph
  {et~al.}(2024{\natexlab{b}})\citenamefont {Sheng}, \citenamefont {Zhao},
  \citenamefont {Li}, \citenamefont {Becattini},\ and\ \citenamefont
  {Hou}}]{Sheng:2024kgg}%
  \BibitemOpen
  \bibfield  {author} {\bibinfo {author} {\bibfnamefont {X.-L.}\ \bibnamefont
  {Sheng}}, \bibinfo {author} {\bibfnamefont {Y.-Q.}\ \bibnamefont {Zhao}},
  \bibinfo {author} {\bibfnamefont {S.-W.}\ \bibnamefont {Li}}, \bibinfo
  {author} {\bibfnamefont {F.}~\bibnamefont {Becattini}},\ and\ \bibinfo
  {author} {\bibfnamefont {D.}~\bibnamefont {Hou}},\ }\bibfield  {title}
  {\bibinfo {title} {{Holographic spin alignment for vector mesons}},\ }\href
  {https://doi.org/10.1103/PhysRevD.110.056047} {\bibfield  {journal} {\bibinfo
   {journal} {Phys. Rev. D}\ }\textbf {\bibinfo {volume} {110}},\ \bibinfo
  {pages} {056047} (\bibinfo {year} {2024}{\natexlab{b}})},\ \Eprint
  {https://arxiv.org/abs/2403.07522} {arXiv:2403.07522 [hep-ph]} \BibitemShut
  {NoStop}%
\bibitem [{\citenamefont {Gomez~Dumm}\ \emph {et~al.}(2023)\citenamefont
  {Gomez~Dumm}, \citenamefont {Noguera},\ and\ \citenamefont
  {Scoccola}}]{GomezDumm:2023owj}%
  \BibitemOpen
  \bibfield  {author} {\bibinfo {author} {\bibfnamefont {D.}~\bibnamefont
  {Gomez~Dumm}}, \bibinfo {author} {\bibfnamefont {S.}~\bibnamefont
  {Noguera}},\ and\ \bibinfo {author} {\bibfnamefont {N.~N.}\ \bibnamefont
  {Scoccola}},\ }\bibfield  {title} {\bibinfo {title} {{Charged meson masses
  under strong magnetic fields: Gauge invariance and Schwinger phases}},\
  }\href {https://doi.org/10.1103/PhysRevD.108.016012} {\bibfield  {journal}
  {\bibinfo  {journal} {Phys. Rev. D}\ }\textbf {\bibinfo {volume} {108}},\
  \bibinfo {pages} {016012} (\bibinfo {year} {2023})},\ \Eprint
  {https://arxiv.org/abs/2306.04128} {arXiv:2306.04128 [hep-ph]} \BibitemShut
  {NoStop}%
\bibitem [{\citenamefont {Mao}\ and\ \citenamefont {Wang}(2017)}]{Mao:2017wmq}%
  \BibitemOpen
  \bibfield  {author} {\bibinfo {author} {\bibfnamefont {S.}~\bibnamefont
  {Mao}}\ and\ \bibinfo {author} {\bibfnamefont {Y.}~\bibnamefont {Wang}},\
  }\bibfield  {title} {\bibinfo {title} {{Effect of discrete quark momenta on
  the Goldstone mode in a magnetic field}},\ }\href
  {https://doi.org/10.1103/PhysRevD.96.034004} {\bibfield  {journal} {\bibinfo
  {journal} {Phys. Rev. D}\ }\textbf {\bibinfo {volume} {96}},\ \bibinfo
  {pages} {034004} (\bibinfo {year} {2017})},\ \Eprint
  {https://arxiv.org/abs/1702.04868} {arXiv:1702.04868 [hep-ph]} \BibitemShut
  {NoStop}%
\bibitem [{\citenamefont {Mao}(2019)}]{Mao:2018dqe}%
  \BibitemOpen
  \bibfield  {author} {\bibinfo {author} {\bibfnamefont {S.}~\bibnamefont
  {Mao}},\ }\bibfield  {title} {\bibinfo {title} {{Pions in magnetic field at
  finite temperature}},\ }\href {https://doi.org/10.1103/PhysRevD.99.056005}
  {\bibfield  {journal} {\bibinfo  {journal} {Phys. Rev. D}\ }\textbf {\bibinfo
  {volume} {99}},\ \bibinfo {pages} {056005} (\bibinfo {year} {2019})},\
  \Eprint {https://arxiv.org/abs/1808.10242} {arXiv:1808.10242 [nucl-th]}
  \BibitemShut {NoStop}%
\bibitem [{\citenamefont {Fu}\ \emph {et~al.}(2025)\citenamefont {Fu},
  \citenamefont {Pawlowski}, \citenamefont {Pisarski}, \citenamefont
  {Rennecke}, \citenamefont {Wen},\ and\ \citenamefont {Yin}}]{Fu:2024rto}%
  \BibitemOpen
  \bibfield  {author} {\bibinfo {author} {\bibfnamefont {W.-j.}\ \bibnamefont
  {Fu}}, \bibinfo {author} {\bibfnamefont {J.~M.}\ \bibnamefont {Pawlowski}},
  \bibinfo {author} {\bibfnamefont {R.~D.}\ \bibnamefont {Pisarski}}, \bibinfo
  {author} {\bibfnamefont {F.}~\bibnamefont {Rennecke}}, \bibinfo {author}
  {\bibfnamefont {R.}~\bibnamefont {Wen}},\ and\ \bibinfo {author}
  {\bibfnamefont {S.}~\bibnamefont {Yin}},\ }\bibfield  {title} {\bibinfo
  {title} {{QCD moat regime and its real-time properties}},\ }\href
  {https://doi.org/10.1103/PhysRevD.111.094026} {\bibfield  {journal} {\bibinfo
   {journal} {Phys. Rev. D}\ }\textbf {\bibinfo {volume} {111}},\ \bibinfo
  {pages} {094026} (\bibinfo {year} {2025})},\ \Eprint
  {https://arxiv.org/abs/2412.15949} {arXiv:2412.15949 [hep-ph]} \BibitemShut
  {NoStop}%
\bibitem [{\citenamefont {Klevansky}(1992)}]{Klevansky:1992qe}%
  \BibitemOpen
  \bibfield  {author} {\bibinfo {author} {\bibfnamefont {S.~P.}\ \bibnamefont
  {Klevansky}},\ }\bibfield  {title} {\bibinfo {title} {{The Nambu-Jona-Lasinio
  model of quantum chromodynamics}},\ }\href
  {https://doi.org/10.1103/RevModPhys.64.649} {\bibfield  {journal} {\bibinfo
  {journal} {Rev. Mod. Phys.}\ }\textbf {\bibinfo {volume} {64}},\ \bibinfo
  {pages} {649} (\bibinfo {year} {1992})}\BibitemShut {NoStop}%
\end{thebibliography}%
\end{document}